\documentclass{vgtc}                          

\ifpdf
  \pdfoutput=1\relax                   
  \usepackage{graphicx}                
  \DeclareGraphicsExtensions{.pdf,.png,.jpg,.jpeg} 
\else
  \usepackage{graphicx}                
  \DeclareGraphicsExtensions{.eps}     
\fi%

\graphicspath{{figures/}{pictures/}{images/}{./}} 

\usepackage{microtype}                 
\PassOptionsToPackage{warn}{textcomp}  
\usepackage{textcomp}                  
\usepackage{mathptmx}                  
\usepackage{times}                     
\usepackage{cite}                      
\usepackage{tabu}                      
\usepackage{booktabs}                  
\usepackage{color}
\usepackage{amssymb}

\onlineid{1754}

\vgtccategory{Research}

\vgtcinsertpkg

\newcommand{\JulienEdit}[1]{\textcolor{black}{#1}}

\title{Topologically Controlled Lossy Compression}

\author{%
Maxime Soler
\thanks{E-mail: soler.maxime@total.com}
\\
\parbox{1.4in}{\scriptsize \centering Total \MaxCor{S.A.}, France.
\\
Sorbonne Universit\'e, CNRS, Laboratoire d'Informatique de Paris 6, F-75005 Paris, France.}
\and M\'elanie Plainchault
\thanks{E-mail: melanie.plainchault@total.com}
\\
\parbox{1.4in}{\scriptsize \centering
Total \MaxCor{S.A.}, France.}
\and Bruno Conche
\thanks{E-mail: bruno.conche@total.com}
\\
\parbox{1.4in}{\scriptsize \centering
Total \MaxCor{S.A.}, France.}
\and Julien Tierny
\thanks{E-mail: julien.tierny@lip6.fr}
\\
\parbox{1.4in}{\scriptsize \centering
Sorbonne Universit\'e, CNRS, Laboratoire d'Informatique de Paris 6, F-75005 Paris, France.}
}

\newcommand{\eqref}[1]{Eq.~\ref{#1}}
\newcommand{\figref}[1]{Fig.~\ref{#1}}
\newcommand{\secref}[1]{Sec.~\ref{#1}}

\newcommand{\domain}{\mathcal{M}}
\newcommand{\range}{\mathbb{R}}

\newcommand{\st}[1]{St(#1)}

\newcommand{\lk}[1]{Lk(#1)}
\newcommand{\lkminus}[1]{Lk^-(#1)}
\newcommand{\lkplus}[1]{Lk^+(#1)}
\newcommand{\sub}[1]{f^{-1}_{-\infty}(#1)}
\newcommand{\sur}[1]{f^{-1}_{+\infty}(#1)}

\newcommand{\offset}{\mathcal{O}}
\newcommand{\persistentDiagram}[1]{\mathcal{D}(#1)}
\newcommand{\Index}{\mathcal{I}}

\newcommand{\bottleneck}{d^B_\infty}
\newcommand{\wasserstein}{d^W_\infty}

\newcommand{\Julien}[1]{\textcolor{black}{#1}}

\newcommand{\Maxime}[1]{\textcolor{black}{#1}}
\newcommand{\MaxCor}[1]{\textcolor{black}{#1}}
\newcommand{\removed}[1]{}

\abstract{
This paper presents a new algorithm for the
lossy compression of scalar data
defined on 2D or 3D regular grids,
with topological control.
Certain
 techniques allow users to control the pointwise error induced by the
compression.
\MaxCor{However, i}n many scenarios \removed{however,} it is desirable to
control in a
similar way the preservation of \removed{more}
high\MaxCor{er}-level notions, such as \removed{the}
\MaxCor{topological} features
\removed{of interest in the data}, in order to provide guarantees
\removed{to the user }
on the outcome of post-hoc data analyses.
This paper presents the first compression technique for scalar data which
supports a \MaxCor{strictly} controlled loss of \removed{the }topological features.
It provides users with
specific guarantees both on the preservation of the important features and on
the size of the smaller features destroyed during compression.
In particular, we present a simple compression strategy based on a
topologically adaptive quantization of the range. Our algorithm provides strong
guarantees on the bottleneck \removed{and Wasserstein }distance\removed{s}
between \removed{the }persistence diagrams of the input and decompressed data\MaxCor{,
specifically those associated with extrema}.
A simple extension of our strategy additionally enables a control on the
pointwise error. We also show how to combine our approach with
state-of-the-art compressors, to further improve the geometrical
reconstruction. Extensive experiments, for comparable compression rates,
demonstrate the superiority of our algorithm in terms of the preservation of
topological features.
We show the utility of our approach by illustrating the compatibility
between the output of post-hoc topological data analysis pipelines, executed
on the input and decompressed data, for simulated or acquired data sets.
We also provide a lightweight VTK-based C++ implementation of our approach for
reproduction purposes.
}

\begin{document}




\firstsection{Introduction}

\maketitle

Data compression is an important tool for the analysis and visualization of
large data sets. In particular, in the context of high performance computing,
current trends and predictions \cite{son14} indicate 
increases of the number of cores \MaxCor{i}n super-computers which evolve faster than
their memory, IO and
network bandwidth\removed{s}. This observation implies that such machines tend to
compute results faster than they are able to store or transfer them.
\JulienEdit{Thus,}
data movement is now recognized as an important bottleneck which
challenges large-scale scientific simulations.
\MaxCor{This}
challenges
even further post-hoc data exploration\removed{s} and \MaxCor{interactive analysis,}
as the output data of \removed{the }simulations often
needs to be transferred to a commodity workstation to conduct such
interactive inspections.
Not only such a transfer is costly in terms of time,
but data can often be too large to fit in the memory of a workstation.
\removed{Thus,}
\MaxCor{I}n this context,
data reduction and compression techniques are needed to
reduce the amount of data to transfer.

While many lossless compression techniques are now well established
\cite{huffman52, ziv77, ziv78, fpzip}, 
scientific data sets
often need to be compressed at more aggressive rates, which requires lossy
techniques \cite{isabela, zfp} (i.e. compression\removed{s} which alter\MaxCor{s}
the data). In the context of \removed{the }post-hoc analysis and visualization of data
which has been compressed with a lossy technique, it is important for users to
understand to what extent their data has been altered\removed{ by the compression}, to
make sure that such an alteration has no impact on the analysis. This motivates
the design of lossy compression techniques with error guarantees.

Several lossy techniques with guarantees have been documented, with a
particular focus on pointwise error \cite{fpzip, sq, sz}. However, pointwise
error is a low level measure and it can be difficult for users to
\Julien{apprehend}
its propagation through their analysis pipeline\Julien{,} and consequently its
impact
on the outcome of their analysis. Therefore, it may be desirable to design
lossy techniques with guarantees on the preservation of
\removed{more }high\MaxCor{er}-level notions,
such as the features of interest in the data.
\Julien{However, the definition of feature\MaxCor{s} \removed{of interest }primarily
depends on the target application, but also on the type of analysis pipeline under
consideration. This motivates, for each possible feature definition, the
design of a corresponding lossy compression strategy with guarantees on the
preservation of the said features. In this work, we introduce a
lossy compression technique that guarantees the preservation of features of
interest, defined with topological notions, hence providing users
with strong guarantees when post-analyzing their data with topological
methods.}


Topological data analysis techniques \cite{edelsbrunner09,
pascucci_topoInVis10, heine16} have 
demonstrated their
ability over the last two decades to capture in a generic, robust and efficient
manner features of interest in scalar data, for many applications:
turbulent combustion
\cite{bremer_tvcg11},
computational fluid dynamics \cite{favelier16},
chemistry \cite{chemistry_vis14},
\removed{or }astrophysics \cite{sousbie11}, etc.
One reason for the success of
topological methods in applications is the possibility for domain experts to
easily translate high level notions into topological terms. For instance, the
cosmic web in astrophysics can be extracted by querying the most persistent
one-dimensional separatrices of the Morse-Smale complex connected to maxima of
matter density \cite{sousbie11}. Many similar feature
definitions in topological terms
can be found in the above application examples.
\Julien{For instance, we detail in \autoref{sec_application} two analysis
pipelines based on topological methods for the segmentation of acquired and
simulated data. In the first case, features of
interest (bones in a medical CT scan) can be extracted as the regions of space
corresponding to the arcs of the split tree \cite{carr00} which are attached to
local maxima of CT intensity. In this scenario, it is important that lossy
compression alters the data in a way that guarantees to preserve
the split tree, to guarantee a faithful segmentation despite compression and
thus, to enable further  measurement, analysis and diagnosis even after
compression.}
Thus, it is \Julien{necessary},
for all
applications involving topological methods in their post-hoc analysis, to
design lossy compression techniques with topological guarantees.

This paper presents, to the best of our knowledge, the first lossy compression
technique for scalar data with such topological guarantees. In particular, we
introduce a simple algorithm based on a topologically adaptive quantization of
the \Maxime{data} range. We carefully study the stability of the persistence diagram
\cite{edelsbrunner02, cohen-steiner05} of the decompressed data \Maxime{compared
to the original one.}
Given a target feature size to preserve, which is
expressed as a persistence threshold $\epsilon$, our algorithm \emph{exactly}
preserves the critical point pairs with persistence greater than $\epsilon$ and
destroys all pairs with smaller persistence. We provide guarantees on the bottleneck
and Wasserstein distances between the persistence diagrams, expressed as a
function of the input parameter $\epsilon$.
Simple extensions to our strategy additionally enable to include a control on
the pointwise error and to combine our algorithm with state-of-the-art
compressors to improve the geometry of the reconstruction, while still
providing strong topological guarantees.
Extensive experiments, for comparable compression rates, demonstrate the
superiority of our technique\removed{s} for the preservation of topological features.
We show the utility of our approach by illustrating the compatibility
between the output of topological analysis pipelines, executed
on the \Maxime{original} 
and decompressed data, for simulated or acquired data
\Julien{(\autoref{sec_application})}.
We also provide a VTK-based C++ implementation
of our approach for reproduction purposes.

\subsection{Related work}
Related existing techniques can be classified into two main
categories, addressing lossless and lossy compression respectively.

Regarding lossless compression, several general purpose algorithms have been
documented, using entropy encoders \cite{HowardV91, Golomb66a,
burrows94, huffman52}, 
dictionaries \cite{ziv77, ziv78} and predictors
\cite{burtscher07, cleary84}. For instance, the compressors associated with the
popular file format \emph{Zip} 
rely on a combination of the LZ77
algorithm \cite{ziv77} and Huffman coding \cite{huffman52}. Such compressors
\Julien{replace}
recurrent bit patterns
in the data
by references to a single
copy of the pattern. Thus, these approaches reach particularly high
compression rates when a high redundancy is present in the data.
Several statistical \cite{IsenburgLS05, fpzip} 
or non-statistical \cite{RatanaworabhanKB06} 
approaches have been proposed for volume
data but
\Maxime{often achieve insufficient compression rates} \Julien{in applications}
\Maxime{(below two \cite{zfp}),}
hence motivating lossy compression techniques.

Regarding lossy compression, many strategies have been documented. Some of them
are now well established and implemented in international standards,
such as GIF\removed{\cite{gif87}} or JPEG\removed{\cite{jpeg92}}.
Such approaches rely for
instance on vector quantization \cite{SchneiderW03} or discrete cosine 
\cite{LauranceM97} \MaxCor{and} related block transforms \cite{zfp}.
However, \Maxime{relatively} 
little work, mostly related to scientific computing applications,
has \Maxime{yet} focused on the definition of lossy compression techniques with an emphasis
on error control, mostly expressed as a bound on the pointwise error. For
instance, \removed{al}though initially introduced for lossless compression, the
\emph{FPZIP} compressor \cite{fpzip} supports truncation of floating point
values, thus providing an explicit relative error control. The \emph{Isabela}
compressor \cite{isabela} supports predictive temporal compression by B-spline
fitting and analysis of quantized error. The fixed rate
compressor \emph{ZFP} \cite{zfp}, based on local block transforms, supports
maximum error control by not ignoring transform coefficients whose effect on the
output is more than a user defined error threshold \cite{LindstromCL16}. More
recently, Di and Cappello \cite{sz} introduced a compressor based on curve
fitting specifically designed for pointwise error control.
\MaxCor{This control i\JulienEdit{s} enforced}\removed{In particular, it error
control}
by explicitly \MaxCor{storing values for which the curve fitting
exceeds the input error tolerance.}
\removed{representing the data values where the
prediction error of the curve fitting exceeds the input error tolerance.}
Iverson et al. \cite{sq} also introduced a compressor, named \emph{SQ},
specifically designed for absolute error control.\removed{The compressor}
\MaxCor{It }supports a
variety of strategies based on range quantization and/or region growing with an
error-based stopping condition. For instance, given an input error tolerance
$\epsilon$, the quantization approach segments the range in contiguous
intervals of width $\epsilon$. Then, the scalar value of each grid vertex is
encoded by the identifier of the interval it projects to in the range. At
decompression, all vertices sharing a unique interval identifier are given a
common scalar value (\removed{typically }the middle of the corresponding interval),
\MaxCor{effectively guaranteeing} a maximum error of $\epsilon$
(for \removed{those }vertices
located in the vicinity of an interval \MaxCor{bound}\removed{extremity}).
Such a range quantization
strategy is particularly appealing for the preservation of topological
features, as one of the key stability results on
persistence diagrams \cite{edelsbrunner02} states that the bottleneck distance
between the diagrams of two scalar functions is bounded by their maximum
pointwise error \cite{cohen-steiner05}. Intuitively, this means that all
critical point pairs with persistence higher than $\epsilon$ in the input will
still be present after a compression based on range quantization. However, a
major drawback of such a strategy is the constant quantization step size, which
implies
that large parts of the
range, possibly devoid of important topological features, will still be
decomposed into contiguous intervals of width $\epsilon$, hence drastically
limiting the compression rate in practice. In contrast, our approach is based on
a topologically adaptive range quantization which precisely addresses
this drawback, enabling superior compression rates.
\removed{Moreover, }\MaxCor{W}e
\MaxCor{additionally} show how to extend our approach with
absolute pointwise error
control\removed{ too}. As detailed in \secref{sec:compressionError}, this strategy
 preserves persistence pairs with persistence larger than
$\epsilon$, \emph{exactly}. In
contrast, since it snaps values to the middle of intervals, simple range
quantization \cite{sq} may alter the persistence of critical point pairs in the
decompressed data, by increasing the persistence of smaller pairs (noise)
and/or decreasing that of larger pairs (features). Such an alteration is
particularly concerning for post-hoc analyses, as it degrades the separation of
noise from features and prevents a reliable post-hoc multi-scale analysis, as
the preservation of the persistence of critical point pairs is no longer
guaranteed. Finally, note that a few approaches also considered topological
aspects \cite{bajaj98, BajajPZ99, taubin98} but for the compression of meshes,
not of scalar data.

\subsection{Contributions}
This paper makes
the following new contributions:
\begin{enumerate}
  \item{\textbf{Approach:} We present the first algorithm for data compression
\MaxCor{specifically designed to enforce topological control.}
\removed{with topological control.} We present a simple strategy and carefully describe
the stability of the persistence diagram of the output data.
In particular, we show that, given a target feature size (i.e. persistence) to
preserve, our approach minimizes both the bottleneck and Wasserstein distances
between the persistence diagrams of the input and \Maxime{de}compressed data.}
  \item{\textbf{Extensions:} We show how this strategy can be easily extended
to additionally include control on the maximum pointwise error. Further, we
show how to combine our compressor with state-of-the-art compressors, to
improve the average error.}
  \item{\textbf{Application:} We present applications of our approach to
\removed{the}post-hoc analys\MaxCor{e}s of simulated and acquired data\removed{sets},
where users can faithfully conduct
advanced topological data analysis on compressed data, with guarantees on the
\Maxime{maximal} size of
missing features and the \emph{exact} preservation of the most
\Maxime{important} ones.}
  \item{\textbf{Implementation:}
  We provide a lightweight VTK-based C++ implementation of
  our approach for reproduction purposes.}
\end{enumerate}


\section{Preliminaries}
This section briefly describes our formal setting and presents an overview of
our approach. An introduction to topological data analysis can be found in
\cite{edelsbrunner09}.

\subsection{Background}
\label{sec:background}

\noindent
\textbf{Input data:}
Without loss of generality, we assume that the input data is a piecewise
linear (PL) scalar field $f : \domain \rightarrow \mathbb{R}$ defined on a
PL $d$-manifold $\domain$ with $d$ equals 2 or 3. It has
value at the
vertices
of $\domain$ and is linearly
interpolated on the simplices
of higher dimension.
Adjacency relations on $\domain$ can be described in a dimension independent
way.
The \emph{star} $\st{v}$ of a vertex $v$ is the set of
simplices
of $\domain$ which contain $v$ as a face. The \emph{link}
$\lk{v}$ is
the set of faces of the simplices of $\st{v}$ which do not intersect
$v$.
The topology of $\domain$
can be described with its
\emph{Betti numbers} $\beta_i$ (the ranks of its homology groups
\cite{edelsbrunner09}), which
correspond in 3D to
the numbers of connected
components ($\beta_0$), non collapsible cycles ($\beta_1$) and voids
($\beta_2$).

\begin{figure}
  \includegraphics[width=\linewidth]{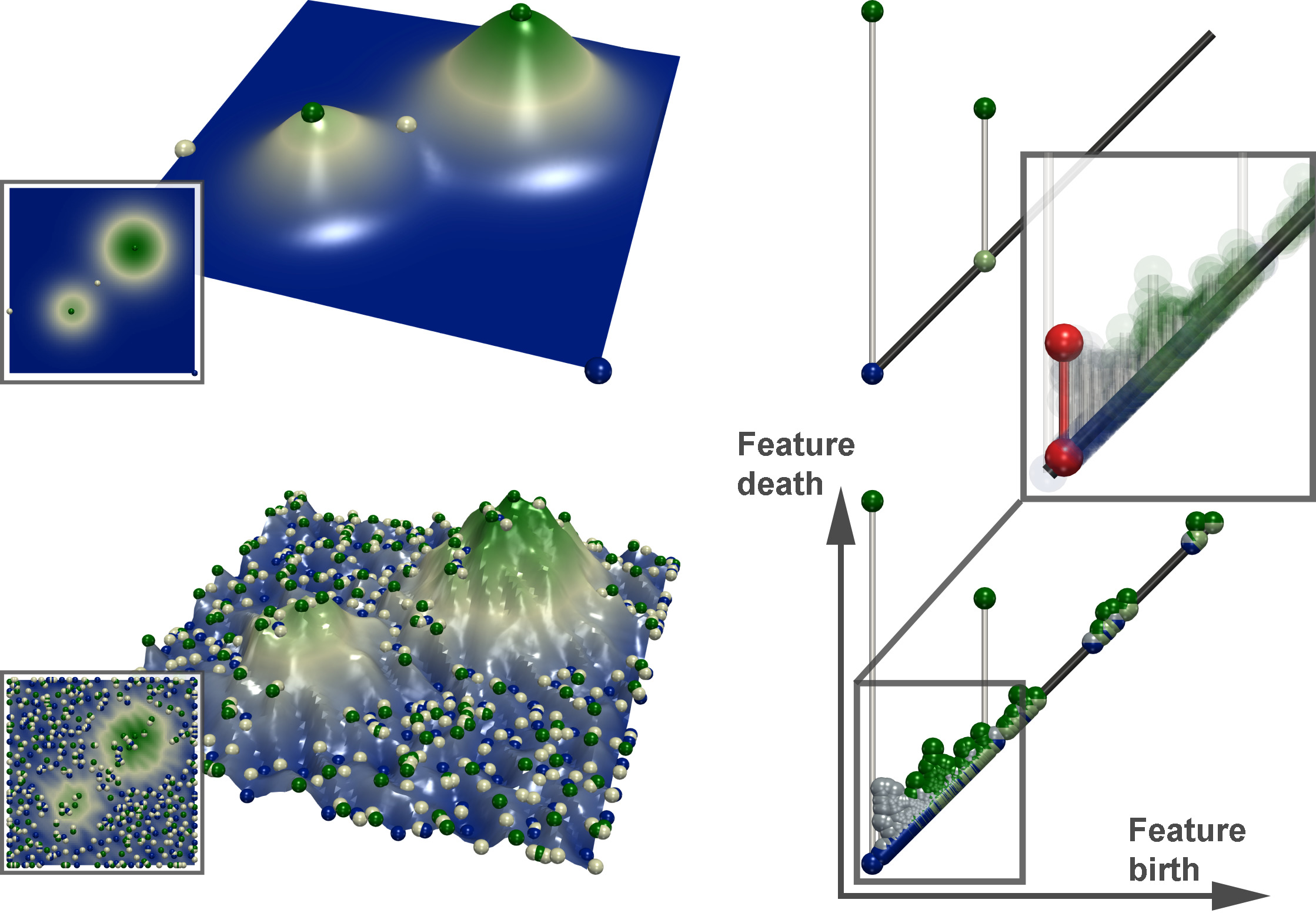}
  \caption{Critical points (spheres, blue: minima, white: saddles, green:
maxima) and persistence diagrams of a clean (top) and noisy (bottom) 2D scalar
field (from blue to green). From left to right: original 2D data, 3D
terrain representation, persistence diagram.
The diagrams clearly exhibit in both cases two large pairs, corresponding to
the two main hills. In the noisy diagram (bottom), small bars near the diagonal
correspond to noisy features in the data.
In this scenario, the bottleneck distance between the diagrams
is \removed{exactly equal to}the persistence
of the largest unmatched feature (red pair in the
zoomed inset, center right)
while the Wasserstein distance is \removed{equal to}the sum
of the persistence of all unmatched pairs.}
  \label{fig:persistence}
\end{figure}

\noindent
\textbf{Critical points:}
For visualization and data analysis purposes, several low-level geometric
features can be defined given the input data. Given an isovalue $i \in \range$,
the
\emph{sub-level set} of $i$,
noted $\sub{i}$, is defined as the pre-image of the open
interval $(-\infty, i)$ onto $\domain$ through $f$:
$\sub{i} = \{ p \in \mathcal{M} ~ | ~ f(p) < i \}$. Symmetrically, the
\emph{sur-level set} $\sur{i}$ is defined by $\sur{i} = \{ p \in \mathcal{M} ~
| ~
f(p) > i \}$. These two objects serve as fundamental
segmentation tools in many analysis tasks \cite{bremer_tvcg11}.
The points of $\domain$ where the Betti numbers of $\sub{i}$ change
are the \emph{critical points} of $f$
(\figref{fig:persistence}) and their associated $f$ values are called
\emph{critical values}. Let $\lkminus{v}$ be
the \emph{lower link} of the vertex $v$: $\lkminus{v} = \{ \sigma \in \lk{v} ~ |
~ \forall u \in \sigma : f(u) < f(v)\}$. The \emph{upper link} $\lkplus{v}$ is
given by $\lkplus{v} = \{ \sigma \in \lk{v} ~ | ~
\forall u \in \sigma : f(u) > f(v)\}$. To classify $\lk{v}$
without ambiguity into either lower or upper links, the restriction of $f$ to
the vertices of $\domain$ is assumed to be injective. This is easily enforced
in practice by a variant of simulation of simplicity \cite{edelsbrunner90}.
This is achieved by considering an associated injective integer offset
$\offset_f(v)$, which initially typically corresponds to the vertex position
offset in memory. Then, when comparing two vertices, if these share the same
value $f$, their order is disambiguated by their offset $\offset_f$. A vertex
$v$
is regular, if and only if both $\lkminus{v}$ and
$\lkplus{v}$ are simply connected. Otherwise, $v$ is a \emph{critical
point} of $f$. Let $d$ be the dimension of $\domain$.
Critical points can be classified with their \emph{index} $\Index$, which
equals 0 for
minima ($\lkminus{v} = \emptyset$), 1 for 1-saddles ($\beta_0(\lkminus{v}) =
2$), $(d - 1)$ for $(d-1)$-saddles ($\beta_0(\lkplus{v}) = 2$) and $d$ for
maxima ($\lkplus{v} =
\emptyset$). Vertices for which $\beta_0(\lkminus{v})$ or
$\beta_0(\lkplus{v})$ are greater than 2 are called \emph{degenerate saddles}.

\noindent
\textbf{Persistence diagrams:}
The distribution of critical points of $f$ can be
represented visually by a topological abstraction called the \emph{persistence
diagram}
\cite{edelsbrunner02, cohen-steiner05} (\figref{fig:persistence}).
By applying the Elder
Rule \cite{edelsbrunner09}, critical points can be arranged in a set of pairs,
such that each critical point appears in only one pair $(c_i, c_j)$ with
$f(c_i) < f(c_j)$ and $\Index(c_i) = \Index(c_j) - 1$.
Intuitively, the Elder Rule \cite{edelsbrunner09} states that if two
topological features of $\sub{i}$ (for instance two connected components) meet
at a given saddle $c_j$ of $f$, the \emph{youngest} of the two features (the
one created last) \emph{dies} at the advantage of the oldest.
For example, if two connected components of $\sub{i}$ merge at a saddle $c_j$,
the one created by the highest minimum $c_i$ (the youngest one) is considered
to die at $c_j$, and $c_i$ and $c_j$ will form a critical point pair.
The persistence diagram
$\persistentDiagram{f}$ embeds each pair $(c_i, c_j)$ in the plane such
that its horizontal coordinate equals $f(c_i)$,
and the vertical coordinate of
$c_i$ and $c_j$ are $f(c_i)$ and $f(c_j)$\Julien{, corresponding respectively
to the \emph{birth} and \emph{death} of the pair}. The height of the pair
$P(c_i, c_j) =
|f(c_j)
- f(c_i)|$ is called the \emph{persistence} and denotes the
life-span of the topological feature created in $c_i$ and destroyed in $c_j$.
\Julien{Thus, features with a short life span (noise) will appear in
$\persistentDiagram{f}$ as low persistence pairs near the diagonal
(\autoref{fig:persistence}, bottom).}
In
low dimension\removed{s}, the persistence of the pairs linking critical points of index
$(0,1)$,
$((d-1),d)$ and $(1,2)$ (in 3D) denotes the life-span of
connected components, voids and non-collapsible
cycles of $\sub{i}$.
In the rest of the paper, when discussing persistence diagrams, we will only
consider critical point pairs of index $(0, 1)$ and $((d-1), d)$. The impact
of this simplification \MaxCor{is}\removed{are}
discussed in \secref{sec:limitations}.
In practice,
persistence diagrams serve as an important visual representation
of the distribution of critical points in a scalar data-set.
Small oscillations in the data due to noise will typically be represented by
critical point pairs with low persistence, in the vicinity of the diagonal.
In contrast, topological features that are the most
prominent in the data will be associated with large vertical bars
(\autoref{fig:persistence}). In many
applications, persistence diagrams help users as a visual guide to
interactively tune simplification thresholds in topology-based, multi-scale 
data segmentation tasks based on the Reeb graph \cite{carr04,
pascucci07, tierny_vis09, gueunet_ldav16,
tierny_vis16}
or the Morse-Smale complex \cite{gyulassy_vis08, gyulassy_vis14}. 

\noindent
\textbf{Distance:}
In order to evaluate the quality of compression algorithms, several metrics
have been defined to evaluate the distance between
the decompressed data, noted $g : \domain
\rightarrow \range$, and the input data, $f : \domain \rightarrow \range$.
The $p$-norm, noted $||f - g||_p$, is a classical example:
\begin{eqnarray}
  ||f - g||_p = \Big(\sum_{v \in \domain} | f(v) - g(v) |^p \Big)
^{{1}\over{p}}
\end{eqnarray}
Typical values of $p$ with practical interests include $p = 2$ and $p
\rightarrow \infty$. In particular, the latter case, called the \emph{maximum
norm}, is used to estimate the maximum pointwise error:
\begin{eqnarray}
  ||f-g||_\infty  = \max_{v \in \domain} | f(v) - g(v)|
\end{eqnarray}
In the compression literature, a popular metric is the
the \emph{Peak Signal to Noise Ratio} (PSNR),
where $|\sigma_0|$ is the number of vertices in $\domain$:
\begin{eqnarray}
  PSNR = 20 ~ log_{10} \Big( {{\sqrt{|\sigma_0|}}\over{2}}
    \times {{\max_{v \in \domain} f(v) - \min_{v \in \domain} f(v)
    }\over{||f - g||_2}}
    \Big)
\end{eqnarray}

\begin{figure*}
  \includegraphics[width=\linewidth]{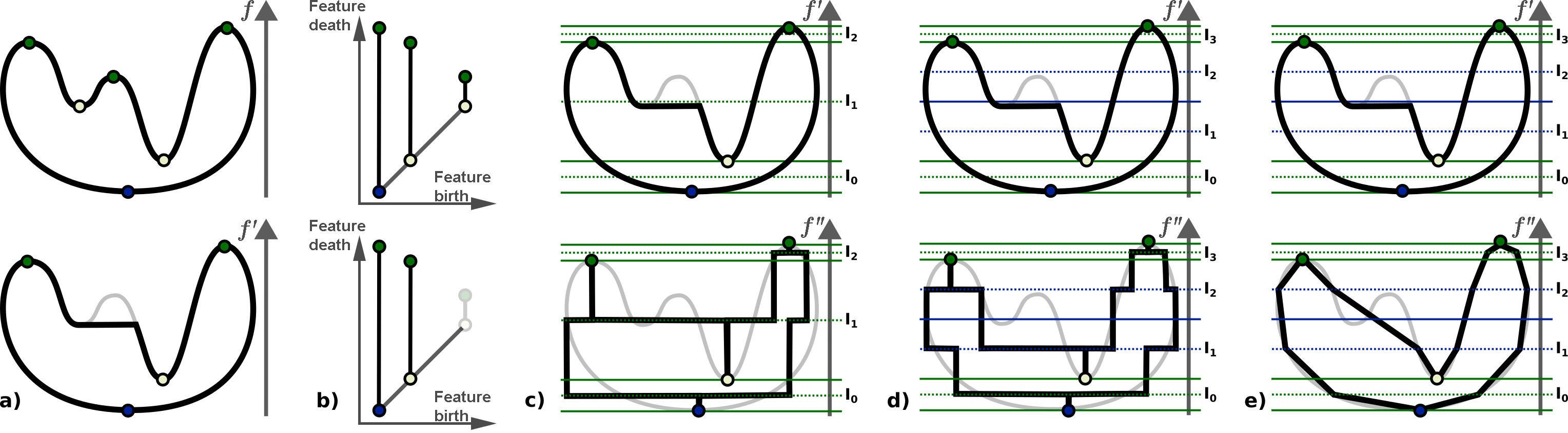}
  \caption{Overview of our topologically controlled lossy compression scheme on
a 2D elevation example. First the input data $f: \domain  \rightarrow \range$
is pre-simplified into a function $f'$ (\Maxime{(a),} 
from top to bottom) to remove all
topological features below a user persistence tolerance $\epsilon$ (as
illustrated by the persistence diagram \Maxime{(b)}). The compression is achieved
by a topologically adaptive quantization of the range, which is segmented along
the critical values of $f'$ \Maxime{(c)}. A quantized function $f''$ is
constructed (\Maxime{(c),}
bottom) to only use a finite set of possible data
values for regular vertices, hence guaranteeing data compression, while still
enforcing original values at critical points. This approach can be extended
with point wise error control (\Maxime{(d)}), by refining each quantization
interval of $f'$ larger than a target width (\Maxime{(d), bottom}).
\Maxime{Moreover,}
our approach
can be combined with any third party compressor \Maxime{(e)} to further
improve the geometry of the compressed data.
}
  \label{fig:overview}
\end{figure*}

In the context of topological data analysis, several metrics
\cite{cohen-steiner05} have been introduced too, in order to 
compare persistence diagrams. In our context, such metrics will be instrumental
to evaluate the preservation of topological features after \Maxime{de}compression. The
\emph{bottleneck} distance \cite{cohen-steiner05}, noted
$\bottleneck\big(\persistentDiagram{f}, \persistentDiagram{g}\big)$, is a
popular
example.
Persistence diagrams can be associated with a pointwise distance, noted
$d_\infty$ inspired by the $\infty$-norm. Given two critical points
$p = (p_x, p_y) \in \persistentDiagram{f}$ and $q = (q_x, q_y) \in
\persistentDiagram{g}$, $d_\infty$ is defined as:
\begin{eqnarray}
  d_\infty (p, q) = \max(|p_x - q_x|, |p_y -q_y|)
\end{eqnarray}
Assume that the number of critical points of all possible index $\Index$
is the same in both
$\persistentDiagram{f}$ and $\persistentDiagram{g}$.
Then the bottleneck
distance $\bottleneck\big(\persistentDiagram{f}, \persistentDiagram{g}\big)$
can be
defined as
follows:
\begin{eqnarray}
  \bottleneck\big(\persistentDiagram{f}, \persistentDiagram{g}\big) =
    \min_{\phi \in \Phi}
      \bigg(
        \max_{p \in \persistentDiagram{f}}
          \Big(
            d_\infty \big(
              p, \phi(p)
            \big)
          \Big)
      \bigg)
\end{eqnarray}
where $\Phi$ is the set of all possible bijections $\phi$ mapping the
critical points of $\persistentDiagram{f}$ to critical points of the same index
$\Index$ in $\persistentDiagram{g}$.
If the numbers of critical points of index $\Index$ do not match in both
diagrams, $\phi$ will be an injection from the smaller set of critical points
to the larger one. Additionally, $\phi$ will collapse the remaining, unmatched,
critical
points in the larger set by mapping each critical point $c_i$ to the other
extremity of its persistence pair $c_j$, in order to still penalize the
presence of unmatched, low persistence features.

Intuitively, in the context of scalar data compression,
the bottleneck distance between two persistence diagrams can be usually
interpreted as the maximal size of the topological features which have not been
maintained through\Maxime{}
compression (\figref{fig:persistence}).

A simple variant of the bottleneck distance, that \MaxCor{is}\removed{will be}
slightly more
informative in the context of data compression, is the \emph{Wasserstein}
distance\MaxCor{,} 
noted $\wasserstein$ (sometimes called
the
\emph{Earth
Mover's Distance} \cite{levina01}), between the persistence
diagrams
$\persistentDiagram{f}$ and $\persistentDiagram{g}$:\removed{, which can be defined as
follows:}
\begin{eqnarray}
  \wasserstein \big(\persistentDiagram{f}, \persistentDiagram{g}\big) =
    \min_{\phi \in \Phi}
      \bigg( \sum_{p \in \persistentDiagram{f}}
        \Big( d_\infty
          \big(
            p, \phi(p)
          \big)
        \Big)
      \bigg)
\end{eqnarray}
In contrast to the bottleneck distance, the Wasserstein distance will take into
account the persistence of \emph{all} the pairs which have not been maintained
through compression (not only the largest one).

\subsection{Overview}
An overview of our compression approach is presented in \figref{fig:overview}.
First, the persistence diagram of the input data $f: \domain \rightarrow
\range$ is computed \MaxCor{so as to evaluate noisy topological features
to later discard. The diagram} consists of \MaxCor{all} critical point pairs of index $(0, 1)$
and $(d-1, d)$. Next, given a target size for the preservation of topological
features, expressed as a persistence threshold $\epsilon$, a simplified
function $f': \domain \rightarrow \range$ is reconstructed \cite{tierny_vis12}
from the persistence diagram of $f$, $\persistentDiagram{f}$, from which all
persistence pairs below $\epsilon$ have been removed (\figref{fig:overview}\Maxime{(a)}).
Next, the
image of $\domain$, $f'(\domain)$, is segmented along each critical value of
$f'$.
A new function $f'' : \domain \rightarrow \range$ is then obtained from $f'$
by assigning to each vertex the mid-value of the interval it maps to. This
constitutes a topologically adaptive quantization of the range
(\figref{fig:overview}\Maxime{(c)}). This
quantization can optionally be further subdivided to enforce a \Maxime{maximal}
pointwise error (\figref{fig:overview}\Maxime{(d)}).
At this point, the
data can be compressed by storing the list of critical values of $f''$ and
storing for each vertex the identifier of the interval it maps to. Optionally,
the input data \MaxCor{$f$}\removed{: \domain \rightarrow \range$}can be compressed independently
by state-of-the-art compressor\MaxCor{s}, such as ZFP \cite{zfp}
(\figref{fig:overview}\Maxime{(e)}).

At decompression, a first function $g': \domain \rightarrow \range$ is
constructed by re-assigning to each vertex the mid-value of the interval it
maps to. Optionally, if the data has been compressed with a third-party
compressor, such as ZFP \cite{zfp}, at decompression,
each vertex value is cropped to the extent of the interval it should map
to. Last, a function $g : \domain \rightarrow \range$ is reconstructed from the
prescribed critical points of $f'$ \cite{tierny_vis12}, to remove any
topological feature resulting from compression artifacts.


\section{Data compression}
\label{sec:dataCompression}
This section presents our topologically controlled compression
scheme. In addition to topological control (\secref{sec:topologicalControl}),
our approach can optionally support pointwise error control
(\secref{sec:errorControl}) as well as combinations with
existing compressors (\secref{sec:zfpCompressor}).
The format of the
files generated by our compressor is described in \secref{sec:fileFormat}.

\subsection{Topological control}
\label{sec:topologicalControl}

The input \Maxime{of} our \Maxime{algorithm}
is the input data, $f : \domain \rightarrow
\range$, as well as the size of the topological features to preserve through
compression. This size is expressed as a persistence threshold $\epsilon$.

First, the persistence diagram of the input data, noted $\persistentDiagram{f}$
is computed. Next, a simplified version of the input data, noted $f' : \domain
\rightarrow \range$, is constructed such that $f'$ admits a persistence
diagram which corresponds to that of $f$, but from which the critical point
pairs with persistence smaller than $\epsilon$ have been removed. This
simplification is achieved by using the algorithm by Tierny and Pascucci
\cite{tierny_vis12}, which iteratively reconstructs sub-level sets to
satisfy topological constraints on the extrema to preserve. In particular,
this algorithm is given as constraints the extrema of $f$ to preserve, which
are in our current setting the critical points involved in
pairs with persistence larger than $\epsilon$. In such a scenario, this
algorithm has been shown to reconstruct a function $f'$ such that
$||f-f'||_\infty \leq \epsilon$ \cite{tierny_vis12}. At this point, $f'$
carries all the necessary topological information that should be preserved
through compression.

In order to compress the data, we adopt a strategy based on range quantization.
By assigning only a small number $n$ of possible data values on the vertices
of $\domain$, only $log_2(n)$ bits should be required in principle for the
storage of each value (instead of 64 for traditional floating point data with
double precision). Moreover, encoding the data with a limited number of
possible values is known to constitute a highly favorable configuration
for post-process lossless compression, which achieves high
compression rates for redundant data.

The difficulty in our context is to define a quantization that respects the
topology of $f'$, as described by its persistence diagram
$\persistentDiagram{f'}$. To do so, we collect all critical values of $f'$ and
segment the image of $\domain$ by $f'$, noted $f'(\domain)$, into a
set of contiguous intervals $I = \{I_0, I_1, \dots
I_n\}$, all delimited by the critical values of $f'$
\Julien{(\autoref{fig:overview}, second column, top)}. Next, we create a new
function $f'' : \domain \rightarrow \range$, where all critical
points of $f'$ are maintained at their corresponding critical value and where
all regular vertices are assigned to the mid-value of the interval $I_i$ they
map to.
This constitutes a topologically
adaptive quantization of the range: only $n$ possible values will be
assigned to regular vertices.
Note that although we modify data values
in the process, the critical vertices of $f'$ are still critical vertices
(with identical indices) in $f''$, as the lower and upper links
(\secref{sec:background}) of each critical point are preserved by construction.

\subsection{Data encoding}
\label{sec:fileFormat}
The function $f''$ is encoded in a two step process. First a topological index
is created. This index stores the identifier of each critical vertex of $f''$
as well as its critical value, and for each of these, the identifier $i$ of the
interval $I_i$ immediately above it if and only if some vertices of $\domain$
indeed project to $I_i$ through $f''$. This strategy enables the save of
identifiers for empty intervals.

The second step of the encoding focuses on \removed{the}data values
of \removed{the}regular
vertices of $f''$. Each vertex of $\domain$ is assigned the identifier $i$ of
the interval $I_i$ it projects to through $f''$.
\MaxCor{
For $n_v$ vertices
and $n_i$ non-empty intervals between $n_c$ critical points, we store
per-vertex interval identifiers ($n_v$ words of $\log_2(n_i)$ bits),
critical point positions in a vertex index ($n_c$ words of $\log_2(n_v)$ bits),
critical types ($n_c$ words of $2$ bits) and critical values ($n_c$ floats).
}

Since it uses a finite set of data values\removed{($n$)}, the buffer storing 
interval assignments for all vertices of $\domain$ is highly redundant.
Thus, we further compress the data (topological index and interval
assignment) with a standard lossless compressor \Julien{(\emph{Bzip2}
\cite{bzip2})}.

\subsection{Pointwise error control}
\label{sec:compressionError}
\label{sec:errorControl}

\begin{figure}
  \includegraphics[width=\linewidth]{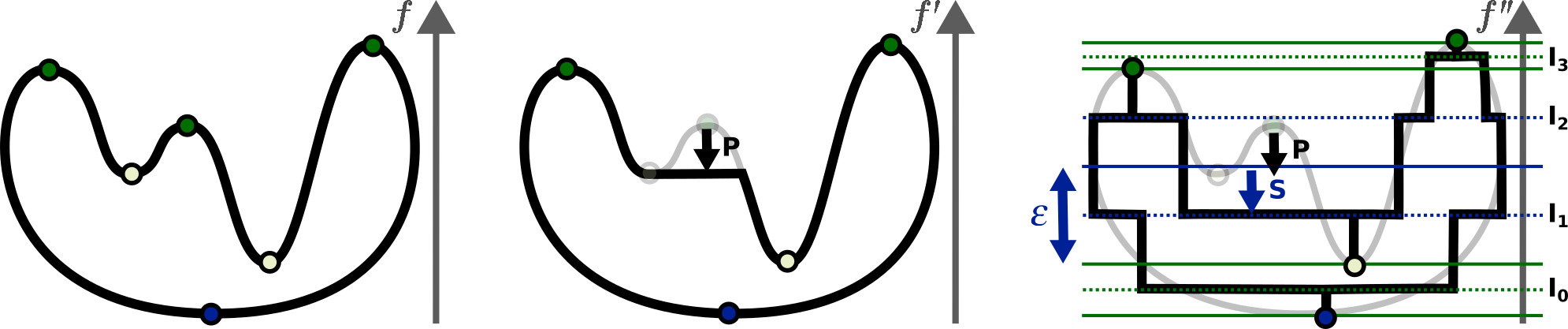}
  \caption{
  Topologically controlled compression with pointwise error control. When
pre-simplifying the input data $f : \domain \rightarrow \range$ (left) into
$f'$ (center), the value variation of each simplified extremum $e$
\removed{is equal to}\MaxCor{equals} the persistence $P$ of the pair it belongs to, which is
bounded by construction by $\epsilon$: $|f(e) - f'(e)| = P \leq \epsilon$
\cite{cohen-steiner05, tierny_vis12}.
When adding pointwise error control, each interval is subdivided such that its
width does not exceed $\epsilon$ (right). Thus, when constructing the quantized
function $f''$ \removed{(}which maps each vertex to the middle of its interval\removed{)},
each simplified extremum $e$ of $f$ may further move by a \emph{snapping} distance
$s$ to the middle of its interval, which is itself bounded by half the width of
the interval ($\epsilon/2$). Thus, \Julien{$|f(e) - f''(e)| = P + s \leq
\epsilon +
\epsilon/2$}.\vspace{-2ex}
  }
  \label{fig:errorControl}
\end{figure}

Our approach has been designed so far to preserve topological
features thanks to a topologically adaptive quantization of the
range. However, this quantization may be composed of
arbitrarily large intervals, which may result in
large pointwise error.

Our strategy can be easily extended to optionally support a maximal pointwise
error with regard to the input data $f : \domain \rightarrow \range$,
still controlled by the parameter $\epsilon$. In particular, this can be
achieved by subdividing each interval (\secref{sec:topologicalControl})
according to a target maximal width $w$, prior to the actual quantization and
data encoding (\secref{sec:fileFormat}). Since the topologically simplified
function $f'$ is guaranteed to be at most $\epsilon$-away from $f$
($||f-f'||_\infty \leq \epsilon$, \secref{sec:topologicalControl}), further
subdividing each interval with a maximum authorized width of $w$ will result in
a maximum error of $\epsilon + w/2$ when quantizing the data into $f''$. For
instance, a local maximum of $f$ of persistence lower than $\epsilon$ can be
pulled down by at most $\epsilon$ when simplifying $f$ into $f'$
\cite{tierny_vis12}
(\figref{fig:errorControl}, center)
and then further pulled down by up to $w/2$ when being
assigned the mid-value
of its new interval (of width $w$, \figref{fig:errorControl}, right). In
practice, for simplicity, we set $w =
\epsilon$. Hence, a maximum pointwise error of $3\epsilon/2$ is guaranteed at
compression ($||f-f''||_\infty \leq 3\epsilon/2$).

\subsection{Combination with state-of-the-art compressors}
\label{sec:zfpCompressor}
The compression approach we presented so far relies on a topologically adaptive
quantization of the range (with optional pointwise error control). The
compression is achieved by only allowing a small number of possible scalar
values in the compressed data, which may typically result in noticeable
\emph{staircase} artifacts. To address this, our method can be optionally
combined seamlessly with any state-of-the-art \Julien{lossy} compressor.
\Julien{For our experiments,} we used \emph{ZFP} \cite{zfp}.
Such a combination is
straightforward at compression time.
\Julien{
In particular, in addition
to the topological index and the compressed quanta identifiers
(\autoref{sec:fileFormat}),
}
the input data $f : \domain \rightarrow
\range$ is \Julien{additionally and independently} compressed
by the third-party compressor (\emph{ZFP}).


\section{Data decompression}
This section describes the decompression procedure of our approach, which is
symmetric to the compression pipeline described in the previous section
(\secref{sec:dataCompression}). This section also further details the
guarantees provided by our approach regarding the bottleneck ($\bottleneck$)
and Wasserstein ($\wasserstein$) distances between the persistence diagrams of
the input data and the decompressed data,
noted $g : \domain \rightarrow \range$ (\secref{sec:guarantees}).

\subsection{Data decoding}
\label{sec:decoding}
First, the compressed data is decompressed with the lossless decompressor
\emph{Bzip2} \cite{bzip2}. Next, a function $g' : \domain \rightarrow \range$
is constructed
based on the topological index and the interval assignment buffer
(\secref{sec:fileFormat}). In particular, each critical vertex is
assigned its critical value (as stored in the topological index) and regular
vertices are assigned the mid-value of the interval they project to, based
on the interval assignment buffer
(\secref{sec:fileFormat}).

\subsection{Combination with state-of-the-art decompressors}
\label{sec:thirdDecompress}
If a state-of-the-art compression method has been used in conjunction with our
approach (\secref{sec:zfpCompressor}), we use
its decompressor to generate the function $g' : \domain \rightarrow \range$.
Next, for each vertex $v$ of $\domain$, if $g'(v)$ is outside of the interval
$I_i$ where $v$ is supposed to project, we snap $g'(v)$ to the closest extremity
of $I_i$. This guarantees that the decompressed data respects
\Julien{
the topological constraints of
$\persistentDiagram{f'}$, as well as}
the optional target pointwise error
(\secref{sec:errorControl}).

\subsection{Topological reconstruction}
\label{sec:finalStep}
The decompressed function $g'$ may contain at this
point extraneous critical point pairs, which were not present in
$\persistentDiagram{f'}$
(\secref{sec:topologicalControl}). For instance, if a state-of-the-art
compressor has been used in conjunction with our approach, arbitrary
oscillations within a given interval $I_i$ can still occur and result in the
apparition of critical point pairs in $\persistentDiagram{g'}$ (with a
persistence smaller than the target interval width $w$,
\secref{sec:errorControl}) which were not present in $\persistentDiagram{f'}$.
The presence of such persistence pairs impacts the distance metrics introduced
in \secref{sec:background}, and therefore impacts the quality of our topology
controlled compression. Thus, such pairs need to be simplified in a
post-process.

Note that, even if no third-party compressor has been used, since our approach
is based on a topologically adaptive quantization of the range, large flat
plateaus will appear in $g'$. Depending on the vertex offset $\offset_{g'} :
\domain \rightarrow \range$ (used to disambiguate flat plateaus,
\secref{sec:background}), arbitrarily small persistence pairs can also occur.
Therefore, for such flat plateaus, $\offset_{g'}$ must be simplified to
guarantee its monotonicity everywhere except at the desired critical vertices
(i.e. those stored in the topological index, \secref{sec:fileFormat}).

Thus, \Maxime{whether a state-of-the-art compressor has been used or not,}
the last step
of our approach consists in reconstructing the function $g :\domain \rightarrow
\range$ from $g'$ by enforcing the critical point constraints of $f'$ (stored
in the topological index) with the algorithm by Tierny and Pascucci
\cite{tierny_vis12}. Note that this algorithm will automatically resolve flat
plateaus, by enforcing the monotonicity of $\offset_g$ everywhere
except at the prescribed critical points \cite{tierny_vis12}. Therefore, the
overall output of our decompression procedure is the scalar field $g : \domain
\rightarrow \range$ as well as its corresponding vertex integer offset
$\offset_g : \domain \rightarrow \mathbb{N}$.

\begin{figure*}
   \includegraphics[width=\linewidth]{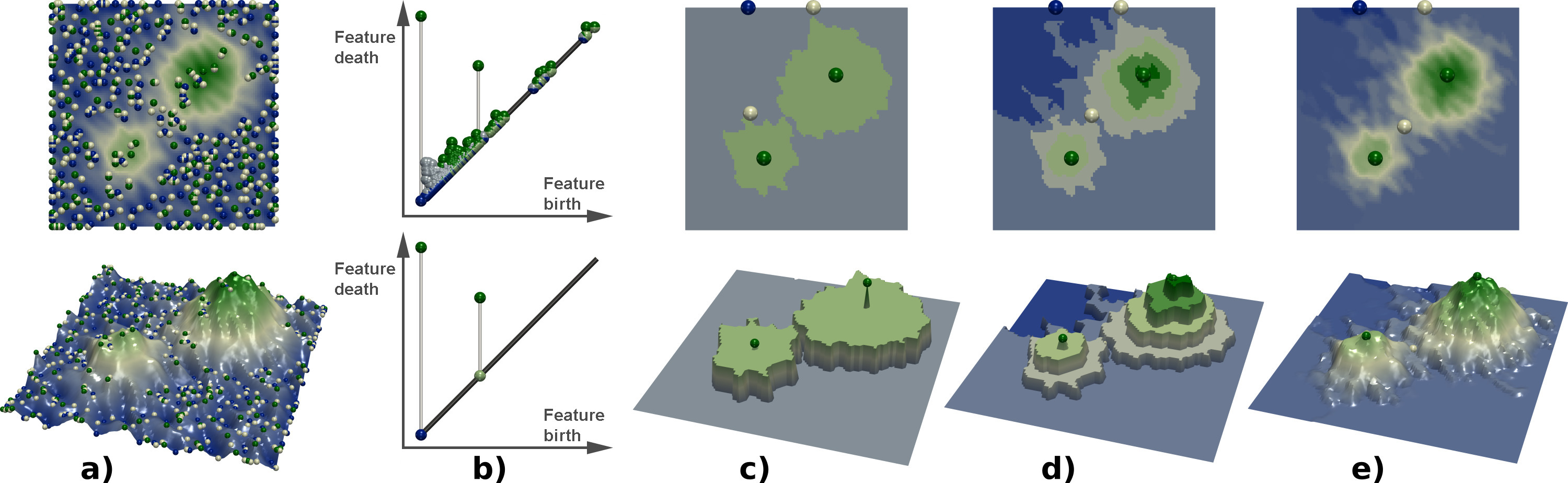}
   \caption{
   Compression of the noisy 2D data set from \figref{fig:persistence}
 (\Julien{(a),} 80,642
 bytes, \Julien{top}: 2D data, \Julien{bottom}: 3D terrain). In all cases
 (\Julien{c-e}),
 our compression algorithm was configured to maintain
 topological features more persistent than $20\%$ of the function range, as
 illustrated with the persistence diagrams \Julien{((b), }
 top: original noisy data $\persistentDiagram{f}$, bottom:
 decompressed data $\persistentDiagram{g}$).
 Our topology controlled compression \Julien{(c)},
 augmented with pointwise error control \Julien{(d)},
 and combined with ZFP \cite{zfp} ((\Julien{e}), one bit per scalar)
 yields compression rates of 163, 50 and 14 respectively.
 \vspace{-2ex}
   }
   \label{fig:compression2D}
 \end{figure*}

\subsection{Topological guarantees}
\label{sec:guarantees}
The last step of our decompression scheme, topological reconstruction
(\secref{sec:finalStep}), guarantees that $\persistentDiagram{g}$ admits no
other critical points
than those of $\persistentDiagram{f'}$ (specified in the topological index).
Moreover, the
corresponding critical values have been strictly enforced
(\secref{sec:decoding}). This guarantees that
$\bottleneck\big(\persistentDiagram{g}, \persistentDiagram{f'}\big) = 0$, and
thus:

\begin{eqnarray}
\begin{array}{l@{}l}
  \bottleneck\big( \persistentDiagram{g}, \persistentDiagram{f} \big) &{}\leq
  \bottleneck\big(\persistentDiagram{g}, \persistentDiagram{f'}\big)  +
  \bottleneck\big( \persistentDiagram{f'}, \persistentDiagram{f} \big)
\\{}&\leq
    \bottleneck\big( \persistentDiagram{f'}, \persistentDiagram{f} \big)
\end{array}
\end{eqnarray}

Since we \removed{have the guarantee}\MaxCor{know}
that $f'$ is $\epsilon$-away from the 
original data $f$ (\secref{sec:topologicalControl})
\Maxime{and due to persistence diagrams stability \cite{cohen-steiner05},}
we then have\removed{the following}\Maxime{:}
\begin{eqnarray}
  \bottleneck\big( \persistentDiagram{g}, \persistentDiagram{f} \big)
    \leq
  ||f - f'||_\infty \leq \epsilon
  \label{eq:guarantee}
\end{eqnarray}
Thus, the bottleneck distance between the persistence diagrams of
the input and
decompressed data is indeed bounded by $\epsilon$, which happens to
precisely describe the size of the topological features that the user wants to
preserve through compression.

Since $\persistentDiagram{f'} \subset \persistentDiagram{f}$ and \removed{that}
$\bottleneck\big(\persistentDiagram{g}, \persistentDiagram{f'}\big) = 0$, we
have $\persistentDiagram{g} \subset \persistentDiagram{f}$. This further
implies that:
\begin{eqnarray}
  \bottleneck\big( \persistentDiagram{g}, \persistentDiagram{f} \big) =
 \max_{(p, q) \in \big(\persistentDiagram{g} \bigtriangleup
\persistentDiagram{f}\big)} P(p, q)
\end{eqnarray}
where $P(p, q)$ denotes the persistence of a critical point pair $(p, q)$ and
where $\persistentDiagram{g} \bigtriangleup \persistentDiagram{f}$ denotes the
symmetric difference between $\persistentDiagram{g}$ and
$\persistentDiagram{f}$ (i.e. the set of pairs present in
$\persistentDiagram{f}$ but not in $\persistentDiagram{g}$). In other words, the
bottleneck distance between the persistence diagrams of the input and
decompressed data \removed{will be}exactly equal\MaxCor{s}\removed{to} the persistence of the most
persistent pair present in $\persistentDiagram{f}$ but not in
$\persistentDiagram{g}$ (\removed{shown}in red in \figref{fig:persistence}). This
guarantees the \emph{exact} preservation of the topological features selected
\removed{by the user }with \MaxCor{an $\epsilon$ persistence threshold.}

\MaxCor{As for} the Wasserstein distance, with the same rationale,
 we \MaxCor{get:}
\begin{eqnarray}
  \wasserstein\big( \persistentDiagram{g}, \persistentDiagram{f} \big)  =
 \sum_{(p, q) \in \big(\persistentDiagram{g} \bigtriangleup
\persistentDiagram{f}\big)} P(p, q)
\end{eqnarray}
In other words, the Wasserstein distance between the persistence diagrams of the
input and decompressed data will be exactly equal to sum of the persistence of
all pairs
present in $\persistentDiagram{f}$ but not in
$\persistentDiagram{g}$ (small bars near the diagonal in
\figref{fig:persistence}, bottom), which corresponds to all the topological
features that the user precisely wanted to discard.

Finally, for completeness, we recall that, if pointwise error control was
enabled, our approach guarantees $||f-g||_\infty \leq 3\epsilon/2$
(\secref{sec:errorControl}).


\section{Results}
This section presents experimental results obtained on a \Julien{desktop
computer with two Xeon CPUs (\Maxime{3.0} GHz, \Maxime{4} cores each), with 64 GB of RAM.}
For the computation of the persistence
diagram and the topological simplification of the data,
we used \Julien{the algorithms by Tierny and Pascucci
\cite{tierny_vis12}
and Gueunet et al. \cite{gueunet_ldav17}, whose implementations are}
available in the Topology ToolKit (TTK) \cite{ttk}. The other components of
our approach (including bottleneck and Wasserstein distance computations) have
been implemented as TTK modules. Note that our approach has been described so
far for triangulations. However, we restrict our experiments to regular grids
in the following as most state-of-the-art compressors (including \emph{ZFP}
\cite{zfp}) have been specifically designed for regular grids. For this,
\removed{purpose,}we \removed{also}use the triangulation data-structure from
TTK,
\removed{as it can}\MaxCor{which} represent\MaxCor{s} implicitly
regular grids with no memory overhead
\MaxCor{using a 6-tet subdivision}.

\figref{fig:compression2D} presents an overview of the compression capabilities
of our approach on a noisy 2D example. A noisy data set is provided on the
input. Given a user threshold on the size of the topological features to
preserve, expressed as a persistence threshold $\epsilon$, our approach
generates decompressed data-sets that all share the same persistence diagram
$\persistentDiagram{g}$ (\Julien{\autoref{fig:compression2D}(b), bottom}),
which is a subset of the diagram
of the input data $\persistentDiagram{f}$
(\Julien{\autoref{fig:compression2D}(b), top})
from which pairs
with a persistence lower than $\epsilon$ have been removed, and those above
$\epsilon$ have been \emph{exactly} preserved. As \Julien{shown} in this
example,
augmenting our approach with pointwise error control or combining it with a
state-of-the-art compressor allows for improved geometrical reconstructions,
\Julien{but} at the expense of much lower compression rates. Note that the
\Julien{\autoref{fig:compression2D}(e)} shows the
result of the compression with \emph{ZFP} \cite{zfp}, which has been augmented
with our topological control. This shows that our approach can enhance any
existing compression scheme, by providing strong topological guarantees on the
output.

\subsection{Compression performance}
\begin{figure}
  \includegraphics[width=0.99\linewidth]{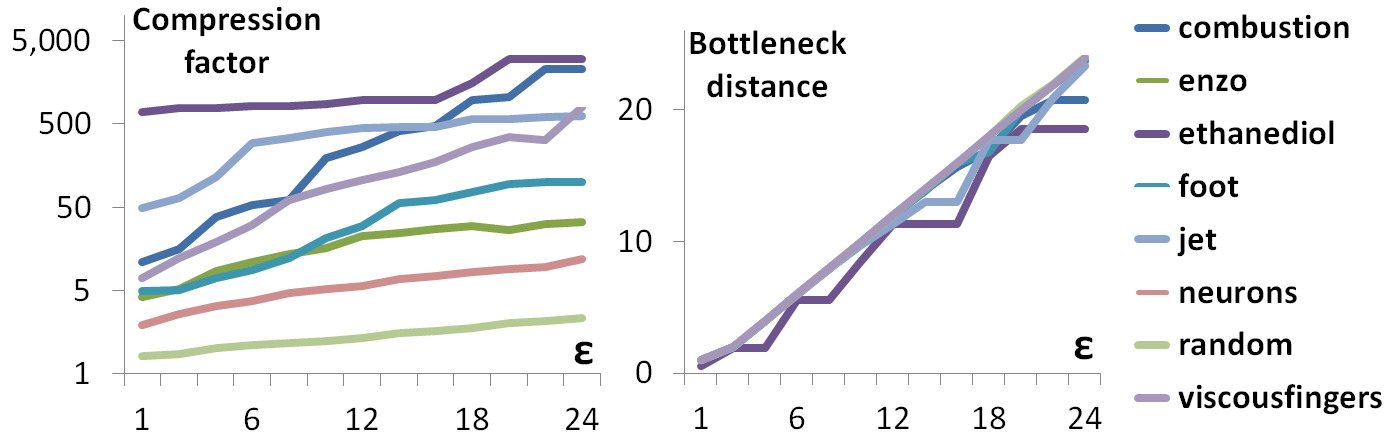}
\hfill
  \caption{
  Performance analysis of our compression scheme (topological control only).
Left: \removed{Evolution of the }\MaxCor{C}ompression rate for various 3D data sets,
\Julien{as a function of the}
target persistence threshold $\epsilon$ (percentage
of the function range). Right: Bottleneck distance
between the persistence diagrams of the input and decompressed data,
$\bottleneck\big(\persistentDiagram{f}, \persistentDiagram{g}\big)$, for
\Julien{increasing} target persistence thresholds $\epsilon$.
\vspace{-2ex}
}
\label{curve:performance}
\end{figure}

We first evaluate the performance of our compression scheme \MaxCor{with topological
control only}
\removed{(without pointwise error control nor combination with third party
compressor)}on a variety of 3D data sets \removed{(synthetic, acquired or simulated
data)} all sampled on \Julien{$512^3$} regular grids. \figref{curve:performance}
(left) presents the evolution of the compression rates
for \Julien{in}creasing target
persistence thresholds $\epsilon$ \removed{(}
expressed as percentages of the \Maxime{data}
range\removed{)}.
This plot confirms that when fine scale structures need to be preserved
(small $\epsilon$ values, \Julien{left}), smaller compression rates are
achieved, while
higher compression rates (\Julien{right}) are reached when \Julien{this}
constraint is relaxed.
Compression rates vary among data sets as the persistence diagrams vary in
complexity.
\MaxCor{The Ethane Diol dataset (topmost curve) is a very smooth function
coming from chemical simulations.}\removed{The elevation data set is a simple elevation function (only one pair
in the diagram, one minimum and one maximum).} High compression \MaxCor{factors}
are achieved for it, \MaxCor{almost} irrespective of $\epsilon$. \removed{A similar behavior is obtained
on the Ethane Diol data set (second top curve), which is a very smooth function
coming from chemical simulations.}\Julien{On the contrary, the random dataset (bottom curve) exhibit\Maxime{s} a complex
persistence diagram, and hence lower compression rates.}

\removed{
There, the input diagram has only few pairs.
Thus, only a small set of $\epsilon$ values will actually modify the
pre-simplified function $f'$, hence the staircase effect in the curve.
On the contrary,
the random data set (bottom curve)
is a synthetic data set where values have been randomly distributed in the
grid, resulting in a complex persistence diagram. There,
the compression rate
also decreases for decreasing $\epsilon$ values.}

In between, all data sets
exhibit the same \Julien{increase in compression rate for increasing
$\epsilon$ values}. Their respective position between the two
extreme configurations of the spectrum (elevation and random) depend on their
input topological complexity (number of pairs in the persistence diagram).

\figref{curve:performance} (right) plots the evolution of the bottleneck
distance between the input and decompressed data,
$\bottleneck\big(\persistentDiagram{f}, \persistentDiagram{g}\big)$, for
\Julien{in}creasing target persistence threshold $\epsilon$, for all data
sets. This
plot shows that all curves are located below the identity diagonal.
This constitutes a practical validation of our guaranteed bound on the
bottleneck distance (\eqref{eq:guarantee}).
\Julien{Note that,} for a given data set, the proximity of its curve to
the diagonal is directly dependent on its topological complexity.
This result confirms the strong
guarantees regarding the preservation of topological features through
compression.

\begin{table}[b]
    \centering
    \caption{\Julien{Detailed computation times on $512^3$
regular grids ($\epsilon=5\%$), with \removed{(left)}and without\removed{(right)}
compression-time simplification. $P$, $S$, $Q$ and $L$ stand for the persistence
diagram, topological simplification, topological quantization and lossless
compression efforts (\%).}
\vspace{-1ex}}
    \label{tab:comparison}
    \label{tab:detailedTimings}
    \scalebox{0.55}{
    \centering
    \begin{tabular}{|l|rrrrrr|rr|r|}
    \toprule
    Data-set
    & \multicolumn{6}{c|}{With compression-time simplification}
     & \multicolumn{2}{c|}{No simplification} &
\multicolumn{1}{c|}{Decompr.} \\
    \cline{2-7}
    \cline{8-9}
    & P (\%) & S (\%) & Q (\%)
      & L (\%) & Total (s) & Compr. Rate
    & Total (s) & Compr. Rate & Time (s)
    \\
    \midrule
    Combustion    & 8.4  & 89.3 & 0.7 & 1.6  & \textbf{593.9}  & 121.3    &
\textbf{64.1}  & 111.1
  & 213.3\\
    Elevation     & 14.6 & 84.1 & 1.2 & 0.1  & \textbf{157.0}  & 174,848.0  &
\textbf{25.3}  &
174,848.0 & 211.5\\
    EthaneDiol    & 12.6 & 86.7 & 0.5 & 0.2  & \textbf{490.0}  & 2,158.6  &
\textbf{63.0}  &
2,158.6 & 228.9\\
    Enzo          & 9.5  & 86.7 & 1.0 & 2.7  & \textbf{695.6}  & 24.5     &
\textbf{91.8}  & 19.8
  & 204.0\\
    Foot          & 12.6 & 81.9 & 1.6 & 3.8  & \textbf{380.6}  & 12.1     &
\textbf{68.4}  & 7.75
  & 205.7\\
    Jet           & 22.2 & 75.7 & 0.6 & 1.5  & \textbf{451.3}  & 315.6    &
\textbf{111.1} & 287.6
  & 220.3\\
    Random        & 15.9 & 76.0 & 2.8 & 5.3  & \textbf{1357.1} & 1.5      &
\textbf{307.7} & 1.5
  & 101.2\\
    \bottomrule
    \end{tabular}
    }
\end{table}

\begin{figure}
\includegraphics[width=\linewidth]{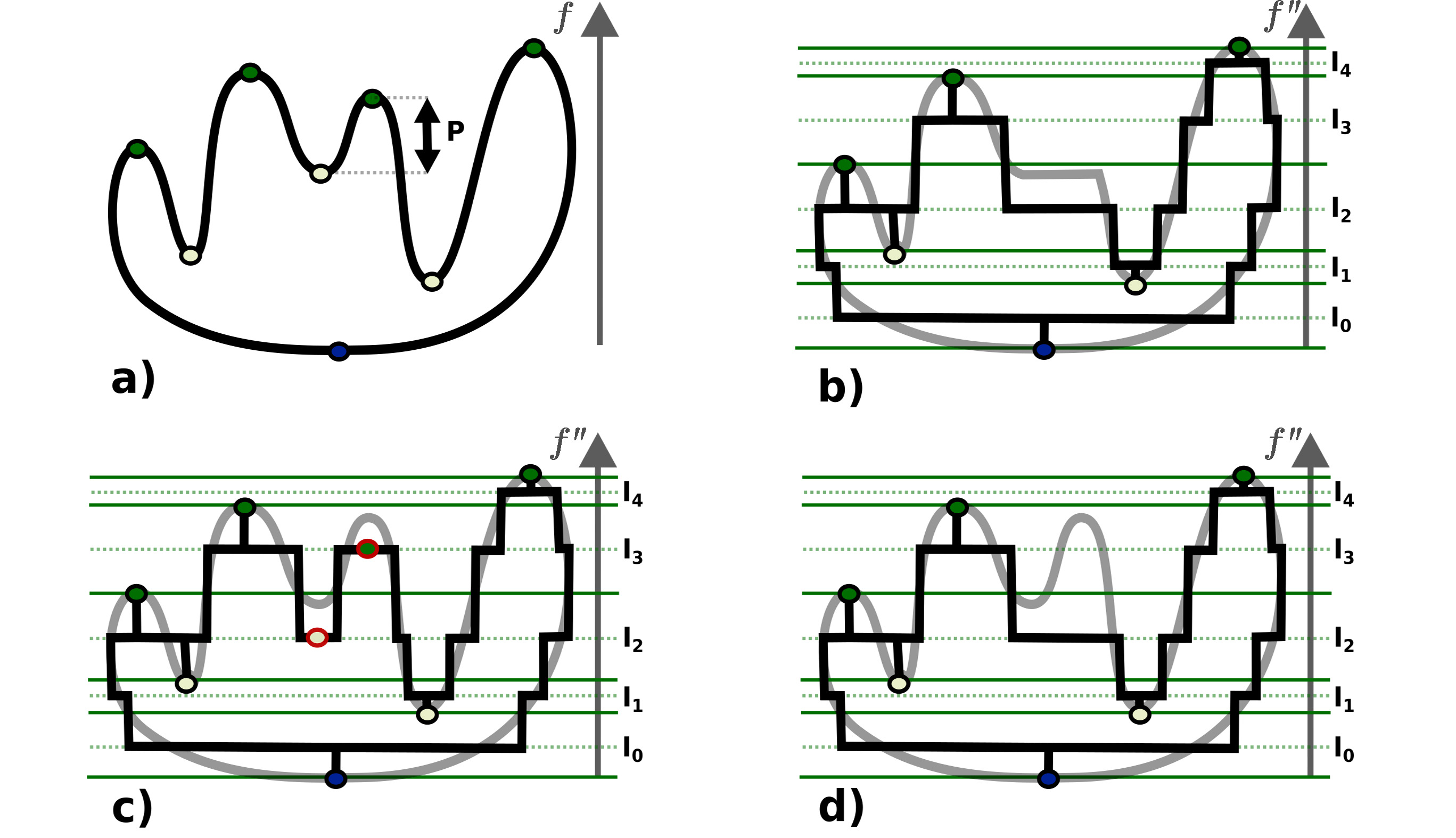}
  \caption{
  \Julien{Topologically controlled compression with (b) and without (c)
  compression-time simplification.
Topological simplification (c to d)
\removed{guarantees to}remove\MaxCor{s} all critical point
pairs not present in the topological index (red circles) and \removed{to}exactly
maintain\MaxCor{s} the others \cite{tierny_vis12}. Thus, simplifying the data only at
decompression (d) \MaxCor{yields}\removed{guarantees} \removed{an}identical decompressed data (d \MaxCor{vs} b).
\removed{However,}\MaxCor{T}he quantized function \MaxCor{then} admits a richer topology (c \MaxCor{vs} b),
which deteriorates compression rates.}
  }
  \label{fig:errorControlNoSimplification}
\end{figure}

\Julien{\autoref{tab:detailedTimings} provides detailed timings for our
approach and shows that most of the compression time (at least 75\%, $S$
column) is spent simplifying the original data $f$ into $f'$
(\autoref{sec:dataCompression}).
If desired, this step can be
skipped to drastically improve time performance, but at the expense of
compression rates (\autoref{tab:detailedTimings}, right). Indeed, as shown in
\autoref{fig:errorControlNoSimplification}, skipping this simplification step
at compression time results in quantized function $f''$ that still admits a
rich topology, and which therefore constitutes a less favorable ground for the
post-process lossless compression (higher entropy).}
Note however\removed{,} that our implementation has not been optimized for
execution time.
\Julien{We leave time performance improvement for future work.}

\subsection{Comparisons}

\begin{figure}
\includegraphics[width=0.99\linewidth]{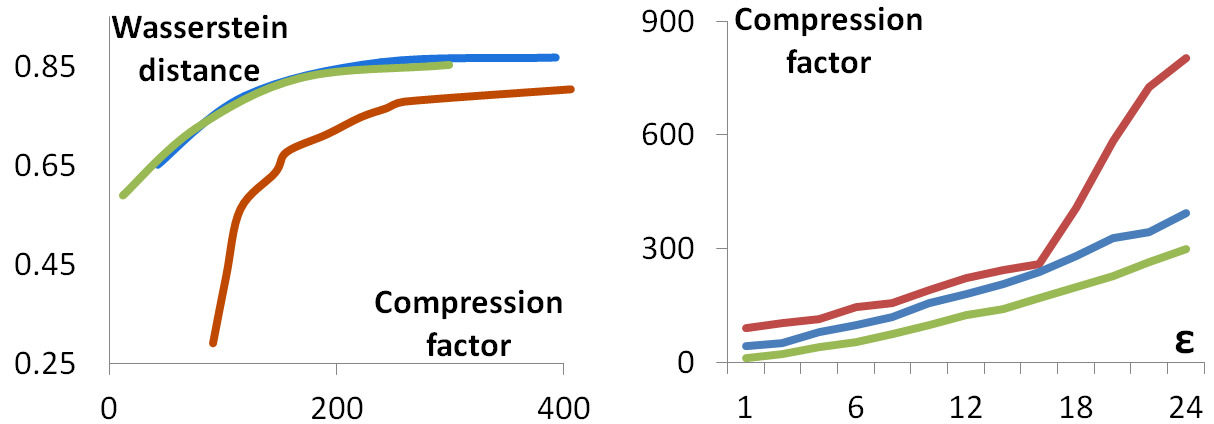}
\hfill
  \caption{Comparison to the \emph{SQ} compressor \cite{sq} (green: SQ-D,
blue: SQ-R, red: our approach). Left:
\removed{Evolution of the }\MaxCor{average n}ormalized
Wasserstein
distance\removed{, averaged over all data sets,}
between the persistence diagrams of the
input and decompressed data,
\removed{
$\wasserstein \big(\persistentDiagram{f},
\persistentDiagram{g}\big)$,
}
for increasing compression rates. Right:
\MaxCor{average c}ompression \MaxCor{factors}\removed{rates, averaged over all data sets,}
for \Julien{in}creasing target
persistence thresholds $\epsilon$.
\vspace{-2ex}
}
  \label{fig:comparisonSQ}
\end{figure}

Next, we compare our approach \MaxCor{with topological control only} to the \emph{SQ} compressor
\cite{sq}, which has been explicitly designed to control
pointwise error.
\Julien{Thus, it is probably}
\Julien{the compression scheme that is the most related to our approach.}
\emph{SQ} proposes two main strategies for data compression, one which is a
straight range quantization (with a constant step size $\epsilon$,
\emph{SQ-R}) and the other which grows regions in the 3D
domain, within a target function interval width $\epsilon$ (\emph{SQ-D}).
Both variants, which we implemented ourselves, provide an
explicit control on the resulting pointwise error ($||f-g||_\infty \leq
\epsilon$). As such, thanks to the stability result on persistence diagrams
\cite{cohen-steiner05}, \emph{SQ} \removed{will}also \MaxCor{bounds the}\removed{guarantee a bounded}
bottleneck distance between the persistence diagrams of the input and
decompressed data. \removed{Indeed, }\MaxCor{E}ach pair completely included within one quantization
step \MaxCor{is indeed}\removed{will be} flattened out. Only the pairs larger than the quantization step
size $\epsilon$ \removed{will}\MaxCor{do} survive through \removed{the}compression. However,
\MaxCor{the latter are snapped to admitted quantization values.}
In practice, this can artificially and arbitrarily reduce the
persistence of certain pairs, and increase \Maxime{the persistence}
of others. This is particularly
problematic as it can reduce the persistence of important features and increase
that of noise, which \removed{will}prevent\MaxCor{s} a reliable multi-scale analysis after
decompression. This difficulty is one of the main motivations which led us to
design our approach.

To evaluate this,
we compare \emph{SQ} to our approach in the light of the
Wasserstein distance between the persistence diagrams of the input and
decompressed data, $\wasserstein \big(\persistentDiagram{f},
\persistentDiagram{g}\big)$. As described in \secref{sec:background}, this
distance is more informative in the context of compression, since
\MaxCor{not only does it track}\removed{it tracks not only} all pairs which have been lost, but \removed{it}also\removed{tracks}
the changes of the
pairs which have been preserved. \figref{fig:comparisonSQ} (left) presents the
evolution of the Wasssertein distance, averaged over all our data sets, for
increasing compression rates. This plot shows that our approach (red curve)
achieves a significantly better preservation of the topological features than
\emph{SQ}, for all compression rates, especially for the lowest ones.
As discussed in the previous paragraph, given a quantization step
$\epsilon$, \emph{SQ} will preserve all pairs more persistent than $\epsilon$
but it will also degrade them, as shown in the above experiment. Another
drawback of \emph{SQ} regarding the preservation of topological features is the
compression rate. Since it uses a constant quantization step size, it may
require many quantization intervals to preserve pairs above a target persistence
$\epsilon$, although large portions of the range may be
devoid of important topological
features. To illustrate this, we compare the compression rates achieved by
\emph{SQ} and our approach, for \Julien{in}creasing values of the parameter
$\epsilon$.
As in \figref{curve:performance}, \Julien{in}creasing slopes can be observed.
However,
our approach always achieves higher compression rates, especially for larger
persistence targets.

\begin{figure}
\vspace{-1ex}
\includegraphics[width=0.99\linewidth
]{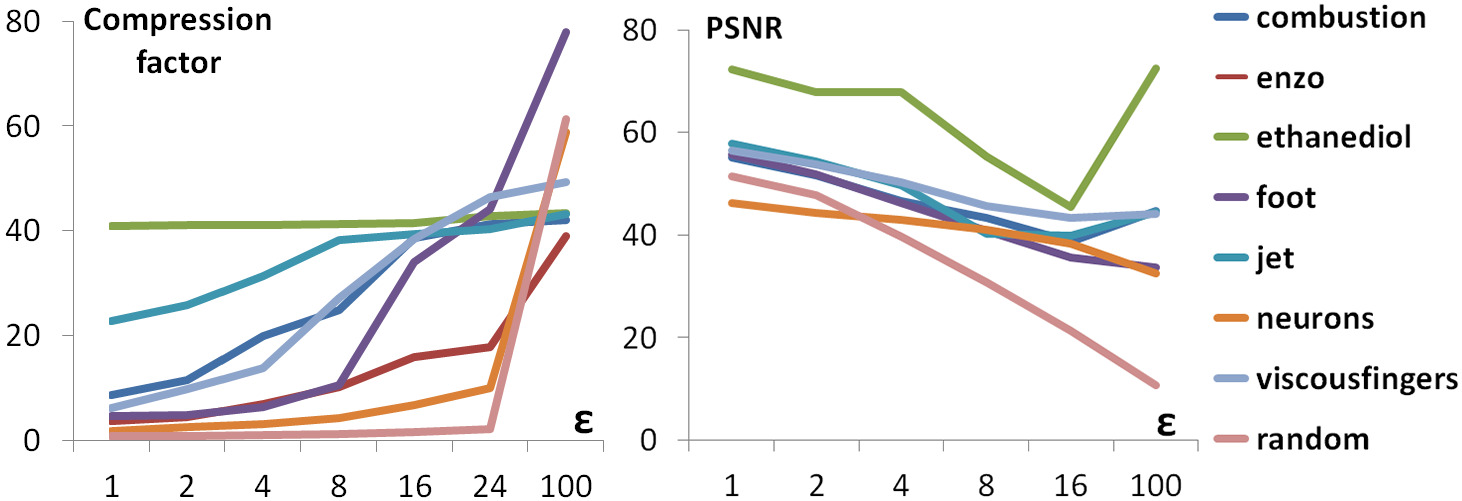}
\hfill
\vspace{-1ex}
  \caption{Augmenting a third party compressor, here \emph{ZFP} \cite{zfp} (1
bit per vertex value),
with topological control. Left: Compression \MaxCor{factors}
\removed{rates}for \Julien{in}creasing
persistence
targets $\epsilon$.
Right: PSNR for \Julien{in}creasing persistence targets $\epsilon$.
In these plots (left and right), a persistence
target of $100\%$ indicates that no topological control was enforced.
\vspace{-2ex}
  }
  \label{curve:zfp}
\end{figure}

Next, we study the capabilities offered by our approach to augment a third
party compressor with topological control (Secs.~\ref{sec:zfpCompressor} and
\ref{sec:thirdDecompress}). In particular, we augmented the \emph{ZFP}
compressor \cite{zfp}, by using the original implementation provided by the
author (with 1 bit per vertex value). \figref{curve:zfp} (left) indicates the
evolution of the compression rates as the target persistence threshold
$\epsilon$ \Julien{in}creases. In particular, in these experiments, a
persistence target
of $100\%$ indicates that no topological control was enforced. Thus,
these curves indicate,
\Julien{apart from this point},
the overhead of topological guarantees over the \emph{ZFP} compressor in terms
of data storage.
These curves, as it could be expected with \figref{curve:performance}, show
that compression rates will rapidly drop down for topologically rich data sets
(such as the random \Julien{one}).
On the contrary, for smoother data sets,
such as
Ethane Diol or Jet, high compression rates can be maintained. This shows that
augmenting a third party compressor with topological control results in
compression rates that adapt to the topological complexity of the input.
\figref{curve:zfp} (right) shows the evolution of the PSNR
(\secref{sec:background}) for decreasing persistence targets $\epsilon$.
Surprisingly, in this context, the enforcement of topological control improves
the quality of the data decompressed with \emph{ZFP}, with higher PSNR values
for \Maxime{little} persistence targets.
This is due to the rather aggressive
compression rate which we used for \emph{ZFP} (1 bit per vertex value) which
tends to add noise to the data. Thanks to our topological control
(\secref{sec:finalStep}), such compression artifacts can be cleaned up by our
approach.

\Julien{
\autoref{tab:zfpCleanup} evaluates the advantage of our topology aware
compression over a standard lossy compression, followed at
decompression by a topological cleanup (which simplifies all pairs
less persistent than $\epsilon$ \cite{tierny_vis12}). In particular, this table
shows that augmenting a third party compressor (such as ZFP) with our
topological control (second line) results in more faithful decompressed data
(lower Wasserstein distances to the original data) than simply executing the
compressor and topologically cleaning the data in a post-process after
decompression (first line). This further motivates our approach for augmenting
existing compressors with topological control.}

\begin{table}[b]
    \centering
    \caption{\Julien{Wasserstein distance between the persistence diagrams of
the original and decompressed data ($\epsilon=1\%$).
    First line: ZFP 1bit/scalar, followed by a topology cleanup procedure.
    Second line: ZFP 1bit/scalar, augmented with our approach.}
    \vspace{-1ex}
    }
    \label{tab:comparison}
    \label{tab:zfpCleanup}
    \scalebox{0.65}{
    \centering
    \begin{tabular}{|l|rrrrrr|}
    \toprule
    Data-set
    & \multicolumn{6}{c|}{$\wasserstein$ } \\
    & Combustion & Elevation & EthaneDiol & Enzo & Foot & Jet \\
    \midrule

    ZFP + Cleanup        & 18.08 & 0.00 & 1.53 & 189.66 & 520,371 &
351.97 \\
    Topology-aware ZFP & 13.73 & 0.00 & 0.40 & 131.11 & 506,714 &
153.45 \\
    \bottomrule
    \end{tabular}
    }
\end{table}

\Julien{Finally, \autoref{tab:timeComparison} provides a comparison of the
running times (for comparable compression rates) between our approach and SQ
and ZFP. ZFP has been designed to achieve high throughput and thus
delivers the best time performances. The running times of our approach are on
par with other approaches enforcing strong guarantees on the decompressed data
(SQ-R and SQ-D).}

\begin{table}[b]
    \centering
    \vspace{-1.5ex}
    \caption{
    Time performance comparison between ZFP, SQ and our approach on
$512^3$ regular grids.
\vspace{-1ex}
    }
    \label{tab:comparison}
    \label{tab:timeComparison}
    \scalebox{0.75}{
    \centering
    \begin{tabular}{|l|rrrr|}
    \toprule
    Data-set
    & \multicolumn{4}{c|}{Time (s).} \\
    \cline{2-5}
    & ZFP & SQ-R & SQ-D  & Ours \\
    \midrule
    Combustion    & 4.6 & 37.6 & 242.9 & 64.1\\
    Elevation     & 7.2& 31.4 & 204.2  & 25.3\\
    EthaneDiol    & 4.7& 34.4 & 197.1  & 63.0\\
    Enzo          & 4.7& 33.0 & 229.5  & 91.8\\
    Foot          & 2.9& 18.2 & 198.0  & 68.4\\
    Jet           & 4.7& 31.4 & 203.4  & 111.1\\
    Random        & 4.1& 31.6 & 182.7  & 307.7\\
    \bottomrule
    \end{tabular}
    }
\end{table}

\subsection{Application to post-hoc topological data analysis}
\label{sec_application}
\begin{figure}
  \includegraphics[width=\linewidth]{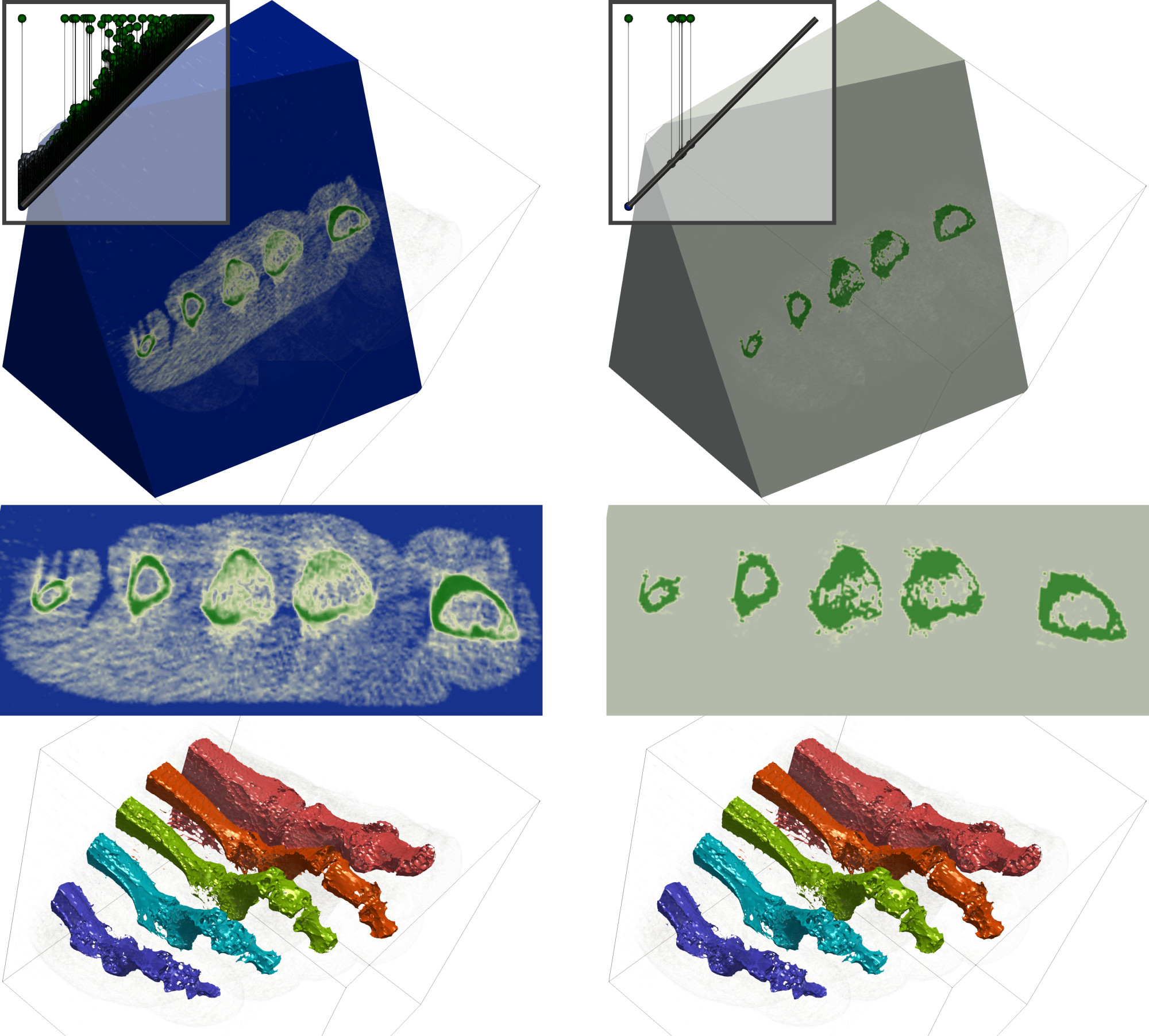}
  \vspace{-4ex}
  \caption{Topology driven data segmentation with multi-scale split trees,
on the original data (left) and the data
\removed{de}compressed with our approach
(right). \removed{From} \JulienEdit{T}op to bottom: persistence diagrams,
sliced views of the data,
output segmentations. 
\MaxCor{The analysis yields compatible outcomes with and without compression,}
as shown with the bottom row, which exhibits identical segmentations
(compression rate: 360).
\vspace{-2ex}
}
  \label{fig:foot}
\end{figure}

A key motivation to our compression scheme is to allow users to faithfully
conduct advanced topological data analysis in a post-process, on the
decompressed data, with guarantees on the compatibility between the outcome of
their analysis and that of the same analysis on the original data.
\Julien{We illustrate this aspect in this sub-section, where all
examples have been compressed by our algorithm, without pointwise
error control nor combination with ZFP.}

We first illustrate this in the context of medical data segmentation with
\figref{fig:foot}, which \MaxCor{shows a foot scan.}\removed{illustrates the
scan of a part of a foot.}
The persistence diagram of the original data counts more than 345 thousands pairs
(top left). The split tree \cite{bremer_tvcg11, carr00} is a
topological abstraction which tracks the connected components of sur-level sets
of the data. It has been shown to excel at segmenting medical data
\cite{carr04}. In this context, users typically compute multi-scale
simplifications of the split tree to extract the most important features. Here,
the user simplified the split tree until it counted only 5 leaves,
corresponding to \removed{the }5 features of interest \removed{in this data }(i.e. the 5 toes).
Next, the segmentation induced by the simplified split tree has been extracted
by considering \removed{the data }regions corresponding to each arc connected to a leaf.
This results immediately in the sharp segmentation of \MaxCor{toe bones.}\removed{the bones of the toes}
(\figref{fig:foot}, bottom left). We applied \removed{exactly the same }\MaxCor{the exact same} analysis pipeline
on the data compressed with our approach. In particular, since it can be known
a priori that this data has only 5 features of interest (5 toes), we compressed
the data with a target persistence $\epsilon$ such that only 5 pairs remained
in the persistence diagram (top right). Although such an aggressive compression
greatly modifies data values, the outcome of the segmentation is identical
\Julien{(\autoref{tab:randIndex})}, for
an approximate compression rate of 360.

\begin{figure}
  \vspace{-1ex}
  \includegraphics[width=0.49\linewidth]{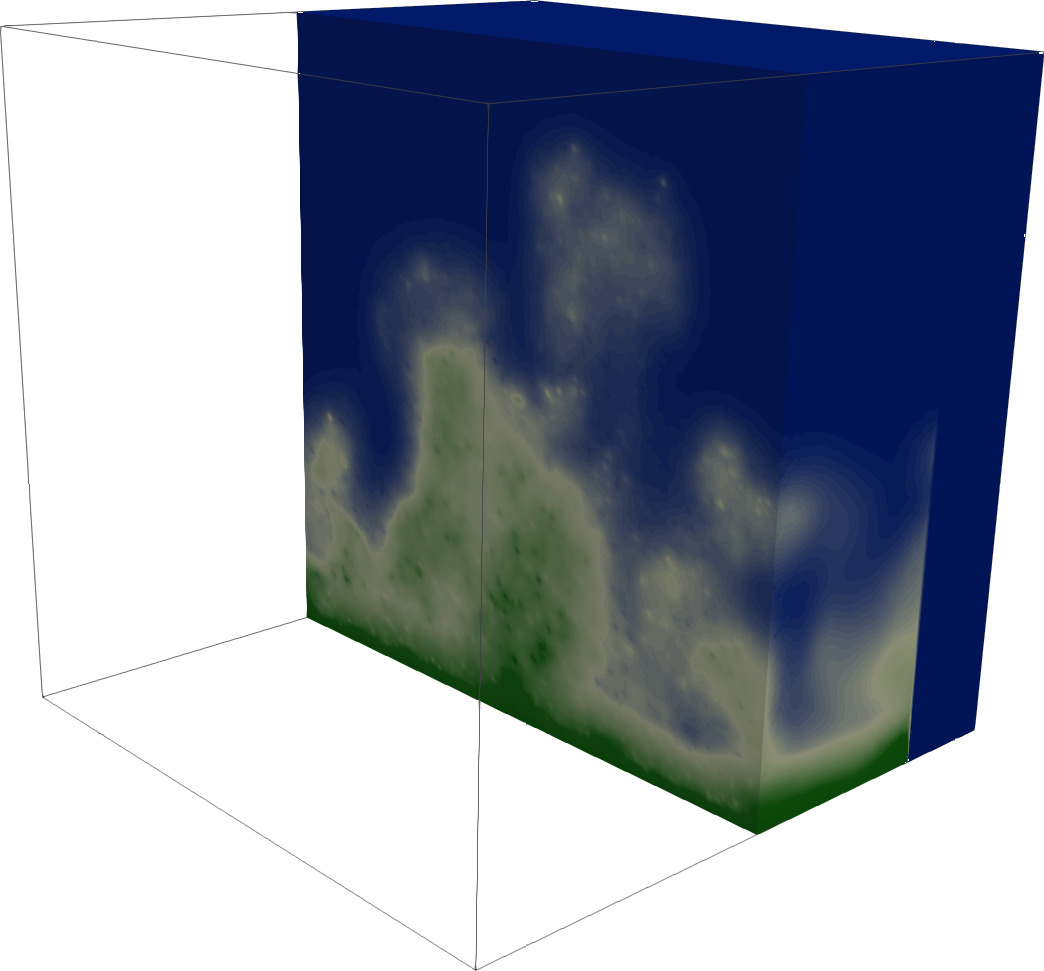}
  \hfill
  \includegraphics[width=0.49\linewidth]{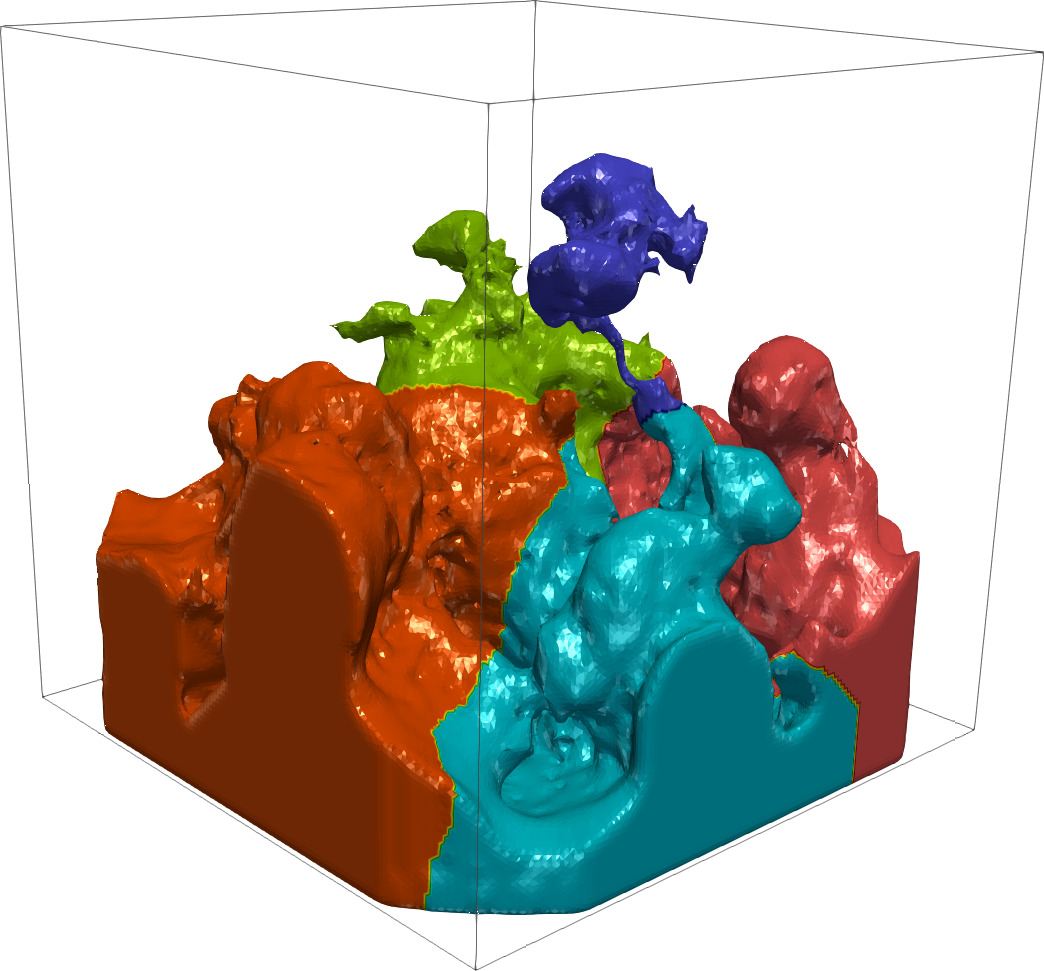}

  ~

\includegraphics[width=0.49\linewidth]{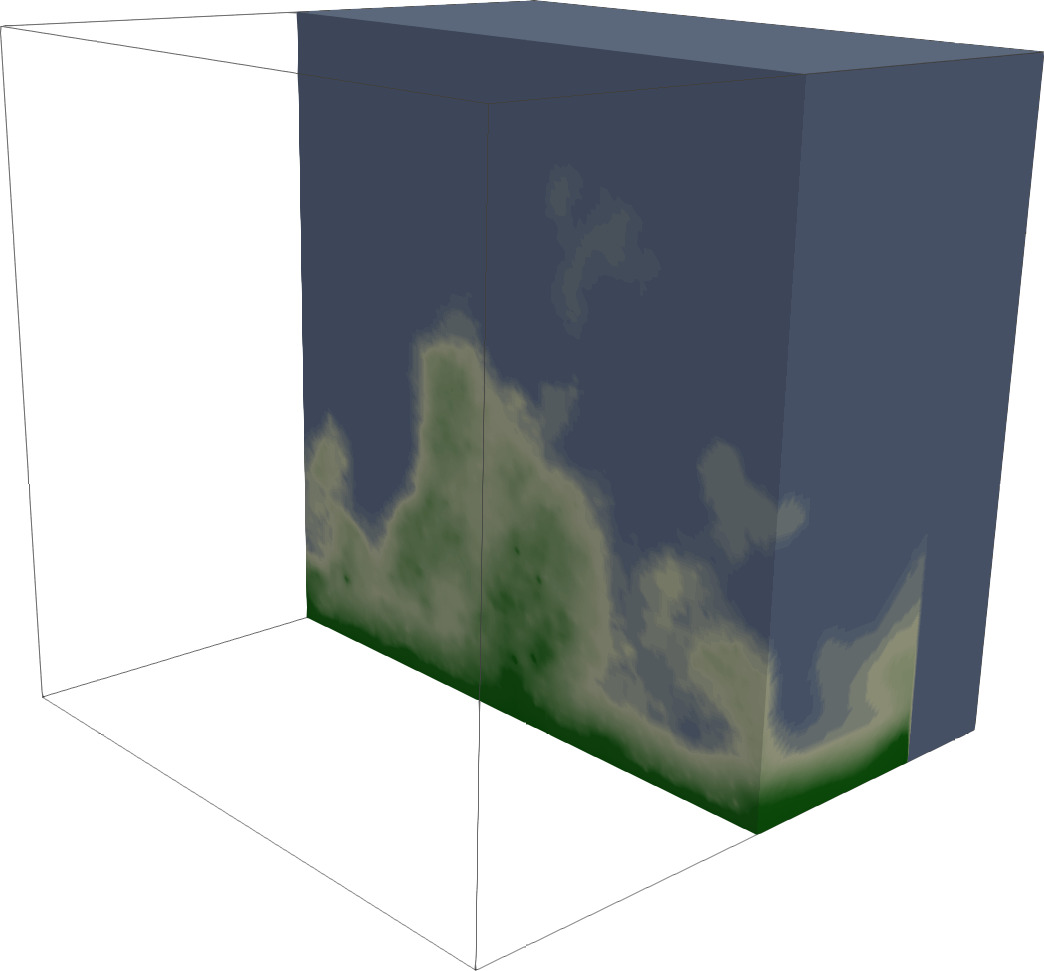}
  \hfill
  \includegraphics[width=0.49\linewidth]{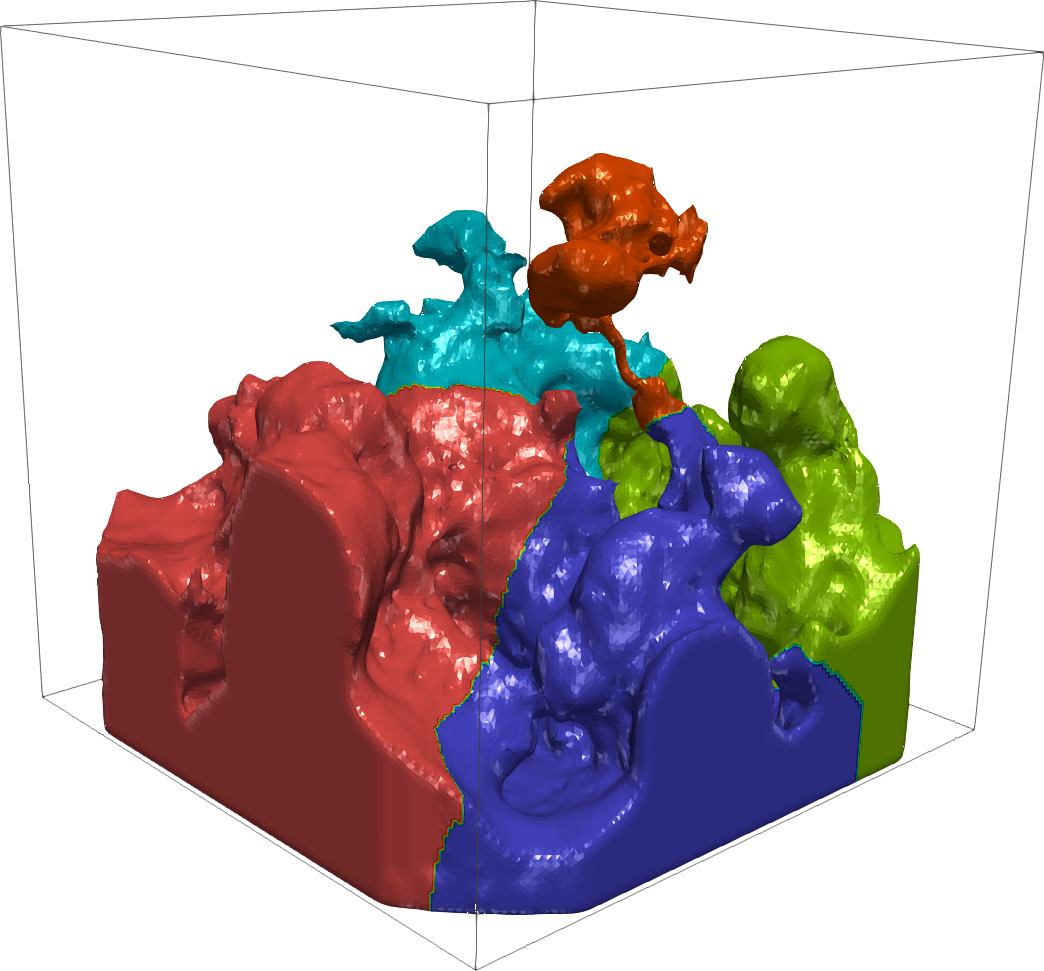}
  \vspace{-2ex}
  \caption{Topology driven data segmentation (right) on a viscous fingering
simulation (left) on the original data (top) and the data
\removed{de}compressed with our approach. \MaxCor{Compatible fingers are extracted after
compression} (compression rate: 56).
  \vspace{-3ex}
  }
  \label{fig:fingers}
\end{figure}

Next, we evaluate our approach on a more challenging pipeline
(\figref{fig:fingers}), where \removed{the }features of interest are not explicitly
related to the persistence diagram of the input data. We consider a
snapshot of a simulation run of viscous fingering and apply the topological data
analysis pipeline described by Favelier et al. \cite{favelier16} for the
extraction and tracking of viscous fingers. This pipeline first isolates the
largest connected component of sur-level set of salt concentration. Next it
considers its height function, on which persistent homology is
applied to retrieve the deepest points of the geometry (corresponding to finger
tips). Finally, a distance field is grown from the finger tips and the Morse
complex of this distance field is computed to isolate \removed{the }fingers.
In contrast
to the previous example, the original data undergoes many transformations and
changes of representation before the \MaxCor{extraction of topological features.}\removed{\Julien{topological} extraction of the
features. } Despite this,
when applied to the data compressed with our scheme, the\removed{outcome of the }
analysis pipeline \removed{is }still
\MaxCor{outputs consistent results with the original data }\removed{compatible with the original data}
(\figref{fig:fingers}, compression rate: 56). Only slight variations can be
perceived in the local geometry of \removed{the }fingers, but
their number \MaxCor{is unchanged}\removed{remains the same} and their overall geometry compatible.
\Julien{\autoref{tab:randIndex} provides a quantitative estimation of the
similarity between \removed{the data }segmentation\MaxCor{s}, before and
after compression \MaxCor{with} several algorithms.
This table shows that our approach enables the
computation of more faithful topological segmentations (higher rand index)
\MaxCor{compared to }\removed{in comparison to other compressors}SQ and ZFP, which further underlines the
superiority of our strategy at preserving topological features.}


\begin{table}[b]
    \centering
    \vspace{-1.5ex}
    \caption{\Julien{Rand index between the outputs of a data segmentation
pipeline based on topological methods (\autoref{sec_application}), before and
after compression, for several methods at compatible compression rates.}
\vspace{-1ex}
}
    \label{tab:comparison}
    \label{tab:randIndex}
    \scalebox{0.85}{
    \centering
    \begin{tabular}{|l|rrrr|}
    \toprule
    Experiment
    & \multicolumn{4}{c|}{Rand index} \\
    & Ours & SQ-R & SQ-D & ZFP \\
    \midrule
    Foot scan       & 1.000 & 0.913 & 0.895 & 0.943\\
    Viscous fingers & 0.985 & 0.977 & 0.977 & 0.973\\
    \bottomrule
    \end{tabular}
    }
\end{table}

\subsection{Limitations}
\label{sec:limitations}
\Julien{Like all lossy techniques, our approach is subject to an input
parameter that controls the \removed{level of }loss, namely the persistence threshold
$\epsilon$ above which features should be strictly preserved. While we believe
this parameter to be intuitive, prior domain knowledge about the size of the
features to preserve may be required. However, conservative values (typically
$5\%$) can be used by default, as they already achieve high compression rates
while preserving most of the features.}
\Julien{In some applications,} ad-hoc metrics \cite{carr04} may be preferred
\Julien{over persistence}. Our
approach can be used in this setting too as the simplification algorithm
that we use \cite{tierny_vis12} supports an arbitrary selection of the critical
points to preserve. However, it becomes more difficult \Julien{then} to
express clear guarantees on the \Julien{compression quality}
in terms of the bottleneck and Wasserstein distances
between the persistence diagrams of the input and decompressed
data.
When \removed{a}pointwise error control is enabled, the $\infty$-norm between the input
and decompressed data is guaranteed by our approach to be bounded by
$3\epsilon/2$. This is due to the topological simplification algorithm that we
employ \cite{tierny_vis12}, which is a flooding-only algorithm. Alternatives
combining flooding and carving \cite{bauer11, tierny_book13} could be
considered to reach a guaranteed $\infty$-norm of $\epsilon$.
Finally, our approach only considers persistence pairs corresponding to
critical points of index $(0, 1)$ and (d-1, d). However $(1, 2)$ pairs may
have a practical interest in certain \MaxCor{3D} applications \removed{in 3D}and it might be
interesting to enforce their preservation \removed{too}through\MaxCor{out}\removed{the} compression. This
would require an efficient \MaxCor{\removed{combinatorial}data reconstruction} algorithm for
\MaxCor{$(1, 2)$ pairs},
which seems challenging \cite{attali13}.


\section{Conclusion}
In this paper, we presented the first compression scheme, to the best of our
knowledge, which provides
strong topological guarantees on the
decompressed data. In particular,
given a target topological feature size
to preserve, expressed as a persistence threshold $\epsilon$, our approach
discards all persistence pairs below $\epsilon$ in order to achieve high
compression rates, while \emph{exactly} preserving persistence pairs above
$\epsilon$. \MaxCor{Guarantees are given}\removed{Our approach provides
guarantees} on \removed{both}the bottleneck and
Wasserstein distances between the persistence diagrams of the
input and
decompressed data. Such guarantees are key to ensure the reliability
of any post-hoc, possibly multi-scale, topological data analysis performed on
\removed{the }decompressed data.
Our approach is simple to implement\MaxCor{;} we provide a lightweight VTK-based C++
\JulienEdit{reference} implementation\removed{of it for reproduction purposes}.

Experiments demonstrated the superiority of our approach in terms of
topological feature preservation in comparison to \Julien{existing}
compressors,
for comparable compression rates. Our approach can be extended to
include pointwise error control. Further, we showed, with the example of the
\emph{ZFP} compressor \cite{zfp}, how to make any third-party compressor
become \emph{topology-aware} by combining it with our approach and making it
benefit from our strong topological guarantees\MaxCor{,
without affecting too much isosurface geometry}. We also showed that, when
aggressive compression rates were selected, our topological approach could
improve existing compressors in terms of PSNR by cleaning up topological
compression artifacts. We finally showed the utility of our approach by
illustrating\Julien{, qualitatively and quantitatively,} the compatibility
between the output of post-hoc topological data analysis pipelines, executed
on the input and decompressed data, for simulated or acquired data sets.
Our contribution enables users to faithfully conduct
advanced topological data analysis on \removed{the }decompressed data,
with guarantees on the
size of \removed{the }missed features and the \emph{exact} preservation of
\removed{the}most prominent ones.

\removed{
\MaxCor{
\subsection{Perspectives}
}}

In the future, we plan to improve \removed{the}practical aspects of our algorithm
\Julien{for \removed{its }in-situ deployment \removed{.}}\MaxCor{and
\JulienEdit{to handle time-varying datasets.}
Runtime limitations will be investigated, with the objective to mitigate
the effects of using (or not) a sequential topological simplification step,
and to determine how many cores are necessary to outperform raw storage.
A streaming version of the algorihm, which would not require the
whole dataset to be loaded at once would be of great interest
in this framework. }\Julien{Finally, }\MaxCor{as} our approach focuses on regular grids,
since, in the case
of unstructured grids, the actual mesh is often the information that requires
the most storage space\MaxCor{, }\removed{. Thus, }we will investigate mesh compression
techniques with guarantees on \MaxCor{topological feature preservation for scalar
fields defined on them.}
\removed{the preservation of the topological features of
scalar fields defined on them.}


\vspace{-.5ex}
\acknowledgments{
\vspace{-1ex}
\small{
This work is partially supported by the Bpifrance grant ``AVIDO'' (Programme
d'Investissements d'Avenir, reference P112017-2661376/DOS0021427) and by the
French National Association for Research and Technology (ANRT), in the framework
of the LIP6 - Total SA CIFRE partnership reference 2016/0010.
The authors would like to thank the anonymous reviewers for their thoughtful
remarks and suggestions.}
}

\bibliographystyle{abbrv-doi}

\bibliography{template}

\begin{thebibliography}{10}

\bibitem{attali13}
D.~Attali, U.~Bauer, O.~Devillers, M.~Glisse, A.~Lieutier.
\newblock Homological reconstruction and simplification in {R}3.
\newblock  {\em SoCG},  117--126, 2013.

\bibitem{bajaj98}
C.~Bajaj D.~Schikore.
\newblock Topology preserving data simplification with error bounds.
\newblock {\em Computers and Graphics}, 22(1):3--12, 1998.

\bibitem{BajajPZ99}
C.~L. Bajaj, V.~Pascucci, G.~Zhuang.
\newblock Single resolution compression of arbitrary triangular meshes with
  properties.
\newblock  {\em IEEE DC},  247--256, 1999.

\bibitem{bauer11}
U.~Bauer, C.~Lange, M.~Wardetzky.
\newblock Optimal topological simplification of discrete functions on surfaces.
\newblock {\em DCG}, 47(2):347--377, 2012.

\bibitem{bremer_tvcg11}
P.~Bremer, G.~Weber, J.~Tierny, V.~Pascucci, M.~Day, J.~Bell.
\newblock Interactive exploration and analysis of large scale simulations using
  topology-based data segmentation.
\newblock {\em IEEE TVCG}, 17(9):1307--1324, 2011.

\bibitem{burrows94}
M.~Burrows D.~Wheeler.
\newblock A block sorting lossless data compression algorithm.
\newblock Technical report, Digital Equipment Corporation, 1994.

\bibitem{burtscher07}
M.~Burtscher P.~Ratanaworabhan.
\newblock High throughput compression of double-precision floating-point data.
\newblock  293--302, 2007.

\bibitem{carr00}
H.~Carr, J.~Snoeyink, U.~Axen.
\newblock Computing contour trees in all dimensions.
\newblock  {\em Symp. on Dis. Alg.},  918--926, 2000.

\bibitem{carr04}
H.~Carr, J.~Snoeyink, M.~van~de Panne.
\newblock Simplifying flexible isosurfaces using local geometric measures.
\newblock  {\em IEEE VIS},  497--504, 2004.

\bibitem{cleary84}
J.~Cleary I.~Witten.
\newblock Data compression using adaptive coding and partial string matching.
\newblock {\em IEEE Trans. Comm.}, 32(4):396--402, 1984.

\bibitem{cohen-steiner05}
D.~Cohen-Steiner, H.~Edelsbrunner, J.~Harer.
\newblock Stability of persistence diagrams.
\newblock  {\em Symp. on Comp. Geom.},  263--271, 2005.

\bibitem{sz}
S.~Di F.~Cappello.
\newblock Fast error-bounded lossy {HPC} data compression with {SZ}.
\newblock  {\em IEEE Symp. on PDP},  730--739, 2016.

\bibitem{edelsbrunner09}
H.~Edelsbrunner J.~Harer.
\newblock {\em Computational Topology: An Introduction}.
\newblock American Mathematical Society, 2009.

\bibitem{edelsbrunner02}
H.~Edelsbrunner, D.~Letscher, A.~Zomorodian.
\newblock Topological persistence and simplification.
\newblock {\em Disc. Compu. Geom.}, 28(4):511--533, 2002.

\bibitem{edelsbrunner90}
H.~Edelsbrunner E.~P. Mucke.
\newblock Simulation of simplicity: a technique to cope with degenerate cases
  in geometric algorithms.
\newblock {\em ToG}, 9(1):66--104, 1990.

\bibitem{favelier16}
G.~Favelier, C.~Gueunet, J.~Tierny.
\newblock Visualizing ensembles of viscous fingers.
\newblock  {\em IEEE SciVis Contest}, 2016.

\bibitem{Golomb66a}
S.~W. Golomb.
\newblock Run-length encodings.
\newblock {\em {IEEE} Trans. on IT}, 12(3):399--401, 1966.

\bibitem{chemistry_vis14}
D.~Guenther, R.~Alvarez-Boto, J.~Contreras-Garcia, J.-P. Piquemal, J.~Tierny.
\newblock Characterizing molecular interactions in chemical systems.
\newblock {\em IEEE TVCG}, 20(12):2476--2485, 2014.

\bibitem{gueunet_ldav16}
C.~Gueunet, P.~Fortin, J.~Jomier, J.~Tierny.
\newblock Contour forests: Fast multi-threaded augmented contour trees.
\newblock  {\em IEEE LDAV},  85--92, 2016.

\bibitem{gueunet_ldav17}
C.~Gueunet, P.~Fortin, J.~Jomier, J.~Tierny.
\newblock Task-based augmented merge trees with {F}ibonacci heaps.
\newblock  {\em IEEE LDAV}, 2017.

\bibitem{gyulassy_vis08}
A.~Gyulassy, P.~T. Bremer, B.~Hamann, V.~Pascucci.
\newblock A practical approach to morse-smale complex computation: Scalability
  and generality.
\newblock {\em IEEE TVCG}, 14(6):1619--1626, 2008.

\bibitem{gyulassy_vis14}
A.~Gyulassy, D.~Guenther, J.~A. Levine, J.~Tierny, V.~Pascucci.
\newblock Conforming morse-smale complexes.
\newblock {\em IEEE TVCG}, 20(12):2595--2603, 2014.

\bibitem{heine16}
C.~Heine, H.~Leitte, M.~Hlawitschka, F.~Iuricich, L.~De~Floriani,
  G.~Scheuermann, H.~Hagen, C.~Garth.
\newblock A survey of topology-based methods in visualization.
\newblock {\em Comp. Grap. For.}, 35(3):643--667, 2016.

\bibitem{HowardV91}
P.~G. Howard J.~S. Vitter.
\newblock Analysis of arithmetic coding for data compression.
\newblock  {\em {IEEE} DCC},  3--12, 1991.

\bibitem{huffman52}
D.~Huffman.
\newblock A method for the construction of minimum-redundancy codes.
\newblock 40(9):1098--1101, 1952.

\bibitem{IsenburgLS05}
M.~Isenburg, P.~Lindstrom, J.~Snoeyink.
\newblock Lossless {C}ompression of {P}redicted {F}loating-{P}oint {G}eometry.
\newblock {\em Computer-Aided Design}, 37(8):869--877, 2005.

\bibitem{sq}
J.~Iverson, C.~Kamath, G.~Karypis.
\newblock Fast and effective lossy compression algorithms for scientific
  datasets.
\newblock  {\em Euro-Par},  843--856, 2012.

\bibitem{isabela}
S.~Lakshminarasimhan, N.~Shah, S.~Ethier, S.~Klasky, R.~Latham, R.~B. Ross,
  N.~F. Samatova.
\newblock Compressing the incompressible with {ISABELA:} in-situ reduction of
  spatio-temporal data.
\newblock  {\em Euro-Par},  366--379, 2011.

\bibitem{LauranceM97}
N.~K. Laurance D.~M. Monro.
\newblock Embedded {DCT} {C}oding with {S}ignificance {M}asking.
\newblock  {\em IEEE ICASPP},  2717--2720, 1997.

\bibitem{levina01}
E.~Levina P.~Bickel.
\newblock The earthmover's distance is the mallows distance: some insights from
  statistics.
\newblock  {\em IEEE ICCV}, vol.~2,  251--256, 2001.

\bibitem{zfp}
P.~Lindstrom.
\newblock Fixed-rate compressed floating-point arrays.
\newblock {\em IEEE TVCG}, 20(12):2674--2683, 2014.

\bibitem{LindstromCL16}
P.~Lindstrom, P.~Chen, E.~Lee.
\newblock Reducing {D}isk {S}torage of {F}ull-3{D} {S}eismic {W}aveform
  {T}omography {(F3DT)} through {L}ossy {O}nline {C}ompression.
\newblock {\em Computers {\&} Geosciences}, 93:45--54, 2016.

\bibitem{fpzip}
P.~Lindstrom M.~Isenburg.
\newblock Fast and efficient compression of floating-point data.
\newblock {\em IEEE TVCG}, 12(5):1245--1250, 2006.

\bibitem{pascucci07}
V.~Pascucci, G.~Scorzelli, P.~T. Bremer, A.~Mascarenhas.
\newblock Robust on-line computation of {R}eeb graphs: simplicity and speed.
\newblock {\em ToG}, 26(3):58, 2007.

\bibitem{pascucci_topoInVis10}
V.~Pascucci, X.~Tricoche, H.~Hagen, J.~Tierny.
\newblock {\em Topological Data Analysis and Visualization: Theory, Algorithms
  and Applications}.
\newblock Springer, 2010.

\bibitem{RatanaworabhanKB06}
P.~Ratanaworabhan, J.~Ke, M.~Burtscher.
\newblock Fast lossless compression of scientific floating-point data.
\newblock  {\em {IEEE} Data Compression},  133--142, 2006.

\bibitem{SchneiderW03}
J.~Schneider R.~Westermann.
\newblock Compression {D}omain {V}olume {R}endering.
\newblock  {\em IEEE VIS},  293--300, 2003.

\bibitem{bzip2}
J.~Seward.
\newblock Bzip2 data compressor.
\newblock \url{http://www.bzip.org}, 2017.

\bibitem{son14}
S.~W. Son, Z.~Chen, W.~Hendrix, A.~Agrawal, W.~k.~Liao, A.~Choudhary.
\newblock Data compression for the exascale computing era - survey.
\newblock {\em Supercomputing Frontiers and Innovations}, 1(2), 2014.

\bibitem{sousbie11}
T.~Sousbie.
\newblock The persistent cosmic web and its filamentary structure: Theory and
  implementations.
\newblock {\em Royal Astronomical Society}, 414(1):350--383, 2011.

\bibitem{taubin98}
G.~Taubin J.~Rossignac.
\newblock Geometric compression through topological surgery.
\newblock {\em ACM Transactions on Graphics}, 17(2):84--115, 1998.

\bibitem{tierny_vis16}
J.~Tierny H.~Carr.
\newblock Jacobi fiber surfaces for bivariate {R}eeb space computation.
\newblock {\em IEEE TVCG}, 23(1):960--969, 2016.

\bibitem{ttk}
J.~Tierny, G.~Favelier, J.~A. Levine, C.~Gueunet, M.~Michaux.
\newblock The {T}opology {T}ool{K}it.
\newblock {\em IEEE TVCG}, 24(1):832--842, 2017.

\bibitem{tierny_book13}
J.~Tierny, D.~Guenther, V.~Pascucci.
\newblock Optimal general simplification of scalar fields on surfaces.
\newblock  {\em Topological and Statistical Methods for Complex Data},  57--71.
  Springer, 2014.

\bibitem{tierny_vis09}
J.~Tierny, A.~Gyulassy, E.~Simon, V.~Pascucci.
\newblock Loop surgery for volumetric meshes: Reeb graphs reduced to contour
  trees.
\newblock {\em IEEE TVCG}, 15(6):1177--1184, 2009.

\bibitem{tierny_vis12}
J.~Tierny V.~Pascucci.
\newblock Generalized topological simplification of scalar fields on surfaces.
\newblock {\em IEEE TVCG}, 18(12):2005--2013, 2012.

\bibitem{ziv77}
J.~Ziv A.~Lempel.
\newblock A universal algorithm for sequential data compression.
\newblock {\em IEEE Trans. on IT}, 23(3):337--343, 1977.

\bibitem{ziv78}
J.~Ziv A.~Lempel.
\newblock Compression of individual sequences via variable-rate coding.
\newblock {\em IEEE Trans. on IT}, 24(5):530--536, 1978.

\end{thebibliography}
\end{document}